\documentclass[%
 aip,
 amsmath,amssymb,
 reprint,%
]{revtex4-2}

\usepackage{graphicx}% Include figure files
\usepackage{dcolumn}% Align table columns on decimal point
\usepackage{bm}% bold math
\usepackage{hyperref}% add hypertext capabilities
\usepackage{tabularx}   % Tabellen, die sich automatisch der 
\usepackage{booktabs}   % Verbesserte Möglichkeiten für Tabellenlayout über horizontale Linien

\usepackage[english]{babel}
\usepackage{siunitx}
\usepackage{upgreek}
\usepackage{subfig}
\usepackage{paralist}
\usepackage[version=4]{mhchem}
\DeclareMathOperator*{\argmax}{arg\,max}

\usepackage{listings}
\usepackage{xcolor}
\usepackage{comment}

\definecolor{codegreen}{rgb}{0,0.6,0}
\definecolor{codegray}{rgb}{0.5,0.5,0.5}
\definecolor{codepurple}{rgb}{0.58,0,0.82}
\definecolor{backcolour}{rgb}{0.95,0.95,0.92}

\lstdefinestyle{mystyle}{
    backgroundcolor=\color{backcolour},   
    commentstyle=\color{codegreen},
    keywordstyle=\color{magenta},
    numberstyle=\tiny\color{codegray},
    stringstyle=\color{codepurple},
    basicstyle=\ttfamily\footnotesize,
    breakatwhitespace=false,         
    breaklines=true,                 
    captionpos=b,                    
    keepspaces=true,                 
    numbers=left,                    
    numbersep=5pt,                  
    showspaces=false,                
    showstringspaces=false,
    showtabs=false,                  
    tabsize=2
}

\let\code\lstinline

\newcommand{\total}{\mathrm d}
\newcommand{\tx}[1]{\text{#1}}

\newenvironment{rcases}
  {\left.\begin{aligned}}
  {\end{aligned}\right\rbrace}

\begin{document}

\preprint{APS/123-QED}

\title{Modeling of a Liquid Leaf Target TNSA Experiment using Particle-In-Cell Simulations and Deep Learning}

\author{B. Schmitz}
 \email{schmitz@temf.tu-darmstadt.de}
    \affiliation{%
    Technische Universität Darmstadt\\
    Institut für Teilchenbeschleunigung und Elektromagnetische Felder (TEMF), Schlossgartenstr.8, 64289 Darmstadt, Germany
    }
\author{D. Kreuter}
 \affiliation{University of Cambridge\\
 Department of Applied Mathematics and Theoretical Physics (DAMTP), Centre for Mathematical Sciences, Wilberforce Road, Cambridge CB3 0WA, United Kingdom}
    \affiliation{%
    Technische Universität Darmstadt\\
    Institut für Teilchenbeschleunigung und Elektromagnetische Felder (TEMF), Schlossgartenstr.8, 64289 Darmstadt, Germany
    }%Lines break automatically or can be forced with \\
\author{O. Boine-Frankenheim}%
    \affiliation{%
     Technische Universität Darmstadt\\
    Institut für Teilchenbeschleunigung und Elektromagnetische Felder (TEMF), Schlossgartenstr.8, 64289 Darmstadt, Germany
    }
    \affiliation{GSI Helmholtzzentrum  f\"{u}r Schwerionenforschung GmbH, Planckstr. 1, 64291 Darmstadt, Germany}

\date{\today}% It is always \today, today,

\begin{abstract}
Liquid leaf targets show promise as high repetition rate targets for laser-based ion acceleration using the Target Normal Sheath Acceleration (TNSA) mechanism and are currently under development. In this work, we discuss the effects of different ion species and investigate how they can be leveraged for use as a possible laser-driven neutron source.
To aid in this research, we develop a surrogate model for liquid leaf target laser-ion acceleration experiments, based on artificial neural networks. The model is trained using data from Particle-In-Cell (PIC) simulations. The fast inference speed of our deep learning model allows us to optimize experimental parameters for maximum ion energy and laser-energy conversion efficiency.
An analysis of parameter influence on our model output, using Sobol' and PAWN indices, provides deeper insights into the laser-plasma system.
\end{abstract}

\keywords{TNSA, Deep Learning, PIC, Liquid Leaf, Lorentz Boost, Multi-Species, Numerical Optimization}%Use showkeys class option if keyword
\maketitle

\section{\label{sec:intro}Introduction}
Laser-accelerated ions have great potential for various applications, such as 
compact medical accelerators \cite{Katayama2022,Eickhoffa,Schardt2010,Linz2007}, neutron sources \cite{Katayama2022,Favalli2019,Kleinschmidt2018, Roth2013} or as injectors for conventional accelerators\cite{Aymar2020}. 
These applications require a high repetition rate to overcome the drawback of the exponential energy distribution, typical for Target Normal Sheath Acceleration.
However, conventional solid-state targets cannot achieve high repetition rates due to engineering difficulties and target supply\cite[chapter 4.2]{zimmer-phd}. 
For this reason, different targets such as gas \cite{Huebl2020} or liquid-based targets \cite{Berglund1998, Wieland2001} are currently being developed.

In this work, we investigate a liquid leaf target \cite{George2019} currently under development at TU Darmstadt.
This target is a major step towards achieving reproducible, high repetition rate ion bunches from laser-plasma interactions, which is necessary for any kind of application.
This new system allows for the operation of a repetitive target with arbitrary \ce{H2O}/\ce{D2O} ratios for the first time, which we investigate in this work.

In particular, we aim to train a surrogate model for a liquid leaf target in a target normal sheath acceleration (TNSA) experiment to understand the characteristics of the liquid leaf and its composition, predict ideal operating points and understand how multiple ion species interact with each other.
The first aim for the target at TU Darmstadt is the creation of a viable compact neutron source, which requires proton energies larger than the production threshold of neutrons ($>\SI{1.7}{\mega\electronvolt}$)\cite{Kleinschmidt2018}.

Previous attempts at modeling laser-plasma acceleration experiments have been made \cite{djordjevic2021,zimmer2021analysis,Ma2021,Djordjevic2021a}. 
With our contributions, presented in this work, we expand on the state-of-the-art by taking more parameters of the experimental setup into account and providing a surrogate for the theory of intricate effects a multi-species target can have on the energy spectrum of the accelerated ions.

Huebl et al.~\cite{Huebl2020} have found a strong influence of the mixture ratio of multiple species on the resulting energy spectra in hydrogen-deuterium targets. 
In \autoref{sec:multi-species}, we expand on their idea and provide indicators for the importance of this effect.

Furthermore, while attempts at modeling a laser-plasma acceleration experiment using neural networks have been made \cite{djordjevic2021}, we extend previous work with a vast number of Particle-In-Cell simulations to train our models. 
The chosen approach via deep learning also ensures that expansion (transfer learning) of the model with experimental data is possible.
We demonstrate our surrogate model's high performance and utility by optimizing an example laser-plasma acceleration experiment, leveraging non-trivial relationships between the experimental parameters not yet understood by theory (see \autoref{sec:applyDL}).

\section{Plasma target models}
The following section details our contributions related to the considered multi-species target experiment as well as the creation of our simulation datasets and the training of our surrogate model.

For this work, we carried out various PIC simulations to generate our datasets.
The bulk of the simulations was computed on the \textit{Virgo} High-Performance Computing cluster~\cite{gsi-virgo} at GSI Helmholtzzentrum, Darmstadt.
For these simulations, we used the \textit{Smilei}~\cite{Derouillat2018} PIC code.
We determined the resulting surrogate by training an artificial neural network from the simulated data.

\subsection{Multi-species target considerations}\label{sec:multi-species}
We are considering a liquid leaf target which consists of
%is made up of 
multiple different atom species.
These ion species differ in their charge-to-mass ratio $q_i/m_i$. 
and an ion can have up to $Z$ different charge states. 
Taking water, for example, one can have up to 8 ionization states of oxygen and an additional one for the hydrogen component. 

Water occurs naturally with different isotopes of hydrogen. Taking into account regular water (\ce{H2O}) and heavy water (\ce{D2O}), an additional degree of freedom—the mixtures between the two—must be considered.
Since several ion species are present, this can be denoted as a $n$-species plasma, where $n$ is the number of ion states in the plasma.

The final non-relativistic kinetic energy of species $i$ accelerated in a constant electric field $E_0$ scales as
\begin{equation}
	E_\text{kin} \propto E_0^2 \frac{q_i^2}{m_i} \; .
\end{equation}
Therefore, species with a higher $q_i^2/m_i$ ratio will gain more energy. Provided that the initial densities are similar, species with higher $q_i^2/m_i$ will deplete most of the available field energy.   
This energy is then split between different particle species and limits the acceleration efficiency of a single species. 

Several species interact with each other, leading to a deformation of the particle spectrum.
Faster particles take electrons from the sheath and screen the acceleration field for the following heavier particles.
These heavier particles are then accelerated in the screened field and hence have less kinetic energy per nucleon and a lower velocity than expected from the assumption above.
Mid-energy lighter particles are accelerated by the following heavier particle front due to the Coulomb force, getting compressed in the momentum space.
This compression causes plateaus and quasi-monoenergetic features to form.

This effect is described in detail and analytically calculated for the asymptotic case for two particle species (deuterium and hydrogen gas) by Huebl et al. \cite{Huebl2020}.
We applied their solutions for 2 species, since we assumed a fully ionized plasma in our simulations to reduce the degrees of freedom inside the plasma.

\begin{figure}
    \centering
    \includegraphics[width=0.95\linewidth]{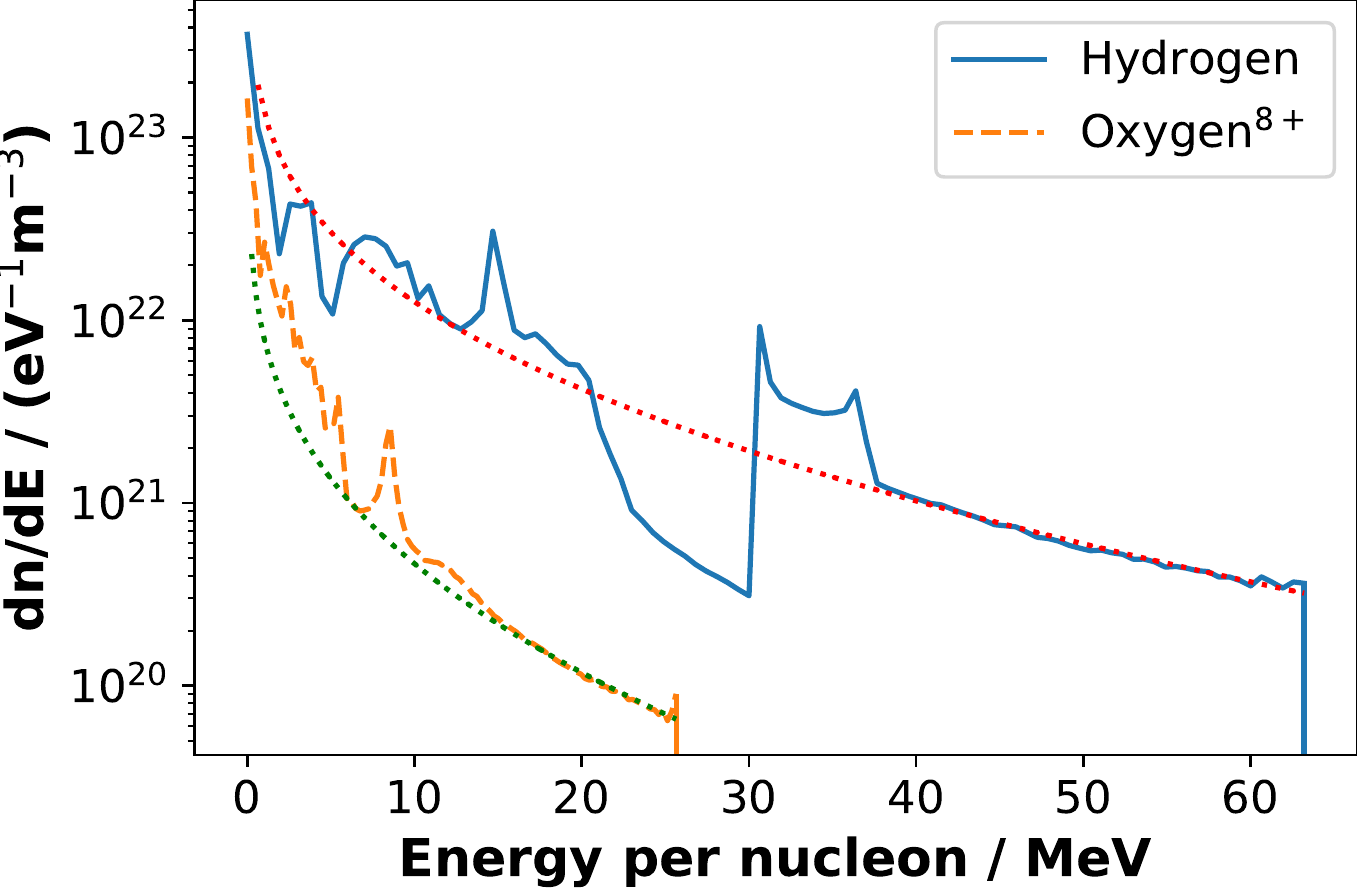}
    \caption{Particle energy spectra of hydrogen and oxygen after TNSA PIC simulation of liquid leaf water target. The dotted lines are the corresponding Mora~\cite{Mora2003} fits for the displayed spectrum.
    Large deviations from the spectrum (\SI{30}{\mega\electronvolt} and up) can be explained by the multi-species effect. The simulation setup is described in \autoref{sec:HPC}.
    The investigated features are sharp and their shape varies. One dimensional simulations have a sharper profile, while higher dimensional ones and real-life experiments are smoother \cite{Huebl2020, Alejo2014}. 
    % Experimental results observed are smoother as well \cite{Alejo2014}.
    }
    \label{fig:picmulti-species}
\end{figure}

The compression effect on the lighter particle spectrum is visualized in \autoref{fig:picmulti-species} using a PIC simulation for regular \ce{H2O}.
Deviations from the ideal Mora\cite{Mora2003, Huebl2020} can be seen. 

We can make two observations:
Firstly, the lower energy part of the spectrum, in this case until half of the maximum energy, is coarser than the corresponding higher energy part of the spectrum.
There is also a peak at around \SI{10}{\mega\electronvolt} in both spectra which makes it possible to compare the spectra against each other.
These peaks are shifted by the same amount as the oxygen cut-off is shifted from the hydrogen plateau, which can be assumed to be a correlation due to the particle interaction in this energy range.\\
Secondly, there is a plateau in the hydrogen spectrum starting at around \SI{30}{\mega\electronvolt}. 
This plateau and the corresponding dip before it deviate fairly strongly from the established Mora theory for TNSA (indicated by the dotted lines).
This drop/increase combination is explainable by the previously introduced multi-species effect and we want to investigate, predict, and leverage this behavior. 

If we can describe and predict this effect, we can optimize our ion beam for specific applications. To do this, we need to find a surrogate model for the full spectrum problem. Further insights gained from considering multiple species become evident in \autoref{sec:examplediscussion}.
Fully ionized oxygen has the same $q_i/m_i$ ratio as deuterium, for example, which reduces the efficiency of deuterium acceleration. In this work, we only modeled the proton part of the spectrum because our data is ambiguous for the Oxygen/Deuterium combination part. However, expanding the dataset to include a sweep of the oxygen's charge number would resolve this ambiguity and yield clearer modeling results for deuterium.

\subsection{Particle-In-Cell Simulations Setup\label{sec:HPC}} 
The simulations reflect a real experiment in reduced dimensions. 
To sample a larger parameter space in a reasonable time, we reduced the dimensions of the simulation to 1.5D.
This means simulating one space and three momentum components. 
The fields are also sampled in three dimensions.

\begin{figure}[h]
    \centering
    \includegraphics[width=0.45\textwidth]{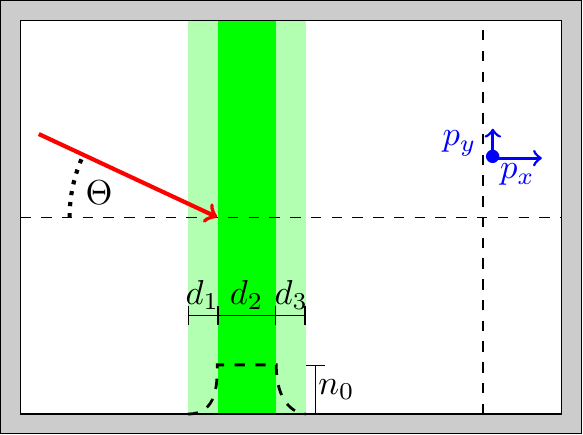}
    \caption{Overview of the simulation setup. Green marks the plasma target. The lighter green areas indicate the pre-plasma and the skirt implemented. 
    The laser, indicated by the red arrow, hits the plasma under an angle $\Theta$ -- relative to the target normal. After the acceleration time, the momenta of the accelerated particles, given in blue, are registered.
    For the liquid leaf target, $d_1$ is assumed to be equal to $d_3$.}
    \label{fig:simSetup}
\end{figure}

We further applied an additional method to account for angle dependency by applying a transverse Lorentz boost to the system. Details on both the Lorentz boost and the method itself can be found in Appendix~\ref{sec:Lorentz}.
A sketch of the full setup is displayed in \autoref{fig:simSetup}.
\subsubsection{Plasma target}
The target in the conducted simulations models a liquid leaf target under development at TU Darmstadt Institute of Nuclear Physics and which is similar to the work by George et al. \cite{George2019}.
The liquid leaf target's width is some cm, while the typical irradiation size of a laser is in the order of \si{\micro\meter}.
We assume, that the surface roughness is negligible and the plasma surface is therefore considered to be planar.

When the target is only dependent on one coordinate, it can be described fully by its particle density profile.
Thus, the simulation only allows movement along the $x$ coordinate and is independent of $y$ and $z$. 
We also assume that the plasma is expanded when the main pulse hits the target, and that the pre-plasma and skirt follow an exponential profile.
We chose the scale length for the exponential profile as \SI{0.4}{\micro\m} so as to be longer than a comparable setup with a cryogenic (i.e. less evaporative) jet target~\cite{Obst2017} while still ensuring a well-defined plasma border. 
The exponential profile thus takes the shape
\begin{align}
	n_\tx{exp}(x) &= \frac{n_0}{1+ \exp(-(x-x_\tx{front})/l_\tx{s})} \label{eq:exppreplasma}
\end{align}
where $l_\tx{s}=\SI{0.4}{\micro\m}$ and $x_\tx{front}$ is the location of the target front. The skirt has identical functional shape for the backside of the target.

Since a liquid leaf target evaporates, we superimposed the typical vapor density distribution for a liquid leaf target, given by
	\begin{align}
		n_\tx{LLT}(r) &\approx n \left(r_\tx{jet}\right) \left(\frac{r_\tx{jet}}{r} \right)^2 \frac{L_\text{L}}{\sqrt{r^2+L_\text{L}^2}} \; , \label{eq:jetpreplasma}
	\end{align}
where $n \left(r_\tx{jet}\right)$ is the water vapor density at the liquid jet surface, $L_\text{L}\approx \SI{3}{\cm}$ is the liquid jet length, and $r_\tx{jet}$ is the liquid jet radius~\cite{Cappa2005}. 
Note that the second term in the above expression has been squared as we expect a faster drop-off of the liquid leaf density in our proposed experimental setup.

The assumed particle densities are $n_0=\SI{6.68e28}{\per\cubic\m}$ and $n\left(r_\text{jet}\right)=\SI{1.62e23}{\per\cubic\m}$ stemming from the liquid density of water and the density estimated at the saturation vapor pressure at \SI{0}{\celsius} \cite{Cappa2005}.
We also introduced a cut-off of the profile \SI{4}{\micro\m} before and after the target, which washes out by approximately \SI{1.4}{\micro\meter} by the time the laser hits the target.
This cut-off is only introduced to optimize the simulation's performance.

We chose to investigate multi-species effects resulting from a combination of different ion species inside the target.
We simulated regular water, heavy water, and a potential mixture between the two.
This mixture is indicated by the mixture parameter listed in \autoref{tab:physical_parameter_list}, which we varied in discrete steps.
The simulation thus consists of up to four species: electrons (e$^-$), hydrogen (H$^+$), deuterium (D$^+$), and oxygen (O$^{n+}$).
As the ionization of oxygen is of importance to the model, this was varied as well.
All particle species follow the same distribution function defined above.

The ion species are initialized cold while the electrons received an initial temperature of \SI{30}{\kilo\electronvolt} to simulate interaction with a pre-pulse. We used Smilei's defaults for particle initialization, including no ionization or radiation model \cite{smilei_namelist_species}.
The length of one cell is the Debye length at the initial electron temperature, around \SI{5}{\nano\meter}. 
For the time resolution, a CFL number of 0.98 was used. 
The interpolation order of the particle shape functions is set to four and the particle per cell count for each species is 800.

\subsubsection{Laserpulse\label{sec:laserpulseparams}}
In a 1.5D simulation, the laser pulse is given by its time profile only. We assumed a Gaussian time profile, using Smilei's \code{tgaussian} profile with the following shape:
	\begin{align}
		I_\tx{envelope}(t) &= \begin{cases} \exp\left( \frac{-(t-\tau_\tx{L})^2}{(\tau_\tx{L}/2)^2/\ln(2)} \right) & \text{if } t \leq 2 \tau_\tx{L}\\
			0 & \text{otherwise} \end{cases} \; , \label{eq:timeEnvelop}
	\end{align}
where $\tau_\tx{L}$ is the main laser pulse duration. In this work, we deal with lasers that have a pulse duration $\tau_\tx{L}<\SI{1}{\pico\second}$ and an $a_0 > 1$.
The laser energy $E_\tx{L}$, pulse length $\tau_\tx{L}$, polarization, incident angle $\theta_\tx{L}$, wavelength $\lambda_\tx{L}$ and the target thickness $d_\tx{T}$ are variable and are uniformly sampled from the defined intervals in \autoref{tab:physical_parameter_list}.
Our thought process in choosing exactly these parameters was that we needed to cover the full system, which required the use of 9 parameters. These parameters were chosen based on two different, sometimes contradictory paradigms: one was to allow the similarity equations to take full effect, while the other was to enable experimental validation of the model (see also Appendix~\ref{sec:unitsAndDimensions} and Appendix~\ref{sec:paramRanges})

\begin{table}[hb]
\caption{Table of the physical parameters that were used for sampling of the input files to the 1.5D PIC simulations. Mixture defines the percentage of hydrogen substituted by deuterium.}
    \begin{tabular}{cccccc}\toprule
		No  &  & Attribute & Sign & Range  & Units \\ \midrule
		1  & Laser  & Energy& $E_\tx{L}$ &[0.001, 50]& \si{\joule} \\
		2  & Laser  & Focus-FWHM  & FWHM & [2,20]   &\si{\micro\meter} \\
		3  & Laser  & Pulse length & $\tau_\tx{L}$ & [15, 150] & \si{\femto\second}\\
		4  & Laser  & Polarization  & &   \{s, p\}   &            \\
		5 & Laser & Incidence angle & $\theta_\tx{L}$&  [0, 85] & \si{\degree}  \\
		6  & Laser  & Wavelength & $\lambda_\tx{L}$ &[550, 1100] & \si{\nano\meter}  \\
		7 & Target & Thickness & $d_\tx{T}$ & [0.6, 3] &\si{\micro\meter} \\
		8 & Target & Mixture & Mix & [0, 100] &\si{\percent} \\
		9 & Target & Oxygen Charge & $Z_\tx{eff}$ & \{7, 8\} & \\ 
		\bottomrule
	\end{tabular}
	\label{tab:physical_parameter_list}
\end{table}

\subsubsection{Simulation Output Quantities\label{sec:simoutput}}
The diagnostics recorded are the particles' $x$-coordinate, all components of the momentum $\vec{p}$, and the macro-particle weight $w$ at the acceleration time 
\begin{align}\label{eq:t_acc}
t_\tx{acc} = \tau_\tx{L} + d_\tx{T}/c_\tx{s} \; ,
\end{align}
where $c_\tx{s}$ is the ion-acoustic velocity. Lécz~\cite{Zsolt2013} has found that this is a suitable acceleration time after which an isothermal plasma expansion model no longer holds. 
From these recorded values we reconstruct the energy spectrum of the particles in the lab frame by using Eq. \eqref{eq:ekindirect}. 
Since all energy spectra have an individual shape and cut-off energy, the spectra were each normalized to the energy range $[0,1]$, counted into 100 bins, and stored as a list together with their respective cut-off energies. 
In order to keep the numbers more practical, we took the logarithm. 
An entry for the results of a simulation thus has the following shape: $\{\ln \left(\frac{\total n}{\total E}\right)_\tx{Bin 1}, \ldots , \ln\left(\frac{\total n}{\total E}\right)_\tx{Bin 100}, E_\tx{max}\}$. 
Exponentiating and re-scaling by $E_\tx{max}$ restores the original energy spectrum accordingly. This same recording scheme is used for all four species for all simulations. 

We chose that the parameters in \autoref{tab:physical_parameter_list} are uniformly sampled with exception of the laser energy $E_\tx{L}$ which we sampled following a square root scale and the mixture was varied in discrete steps.
This type of sampling results in significantly more simulations with low $a_0$ than with high $a_0$. 
To deal with this we forced additional simulations onto dedicated intervals of $a_0$.
Although the laser focus-FWHM is technically not relevant in the 1D case we sampled it nonetheless such that together with the sampled laser energy and pulse length the correct $a_0$ was written in the input file.
This also ensures comparability with higher-order simulations and experimental data.

\subsubsection{Simulation statistics}
We used the setup described above to create a dataset of simulations for our subsequent surrogate model. 
All parameters were stochastically sampled and their combination can be thought of as a sparse grid. The Virgo cluster employs the Simple Linux Utility for Resource Management (SLURM)~\cite{yoo2003slurm} to schedule incoming jobs where up to \num{10000} jobs can be added to the queue simultaneously. The jobs were queued using a script to sample a certain number of parameter combinations and then start a simulation job for each of them.
The number of simulations varies between the different species.
There were \num{508200} simulations for hydrogen and \num{762426} simulations for deuterium, resulting in a total of \num{1270626} simulations. However, the precise number of simulations is not crucial, as long as the number of simulations is in a similar order of magnitude, the results should be comparable.
The reduced model, which utilizes only the pure \ce{H2O} data without \ce{D2O} component was trained on a subset of the full dataset with \num{68973} entries accordingly.

\subsubsection{Limitations of 1.5D PIC}
We used 1.5D simulations as mentioned earlier.
These low-dimensional simulations do have some drawbacks. 
While they, together with our introduced transversal Lorentz boost method (Appendix \ref{sec:Lorentz}), are capable of describing several effects, some are not possible.
The main limitation is created by the expansion of the plasma behind the target. 
In one spatial dimension, no transversal drift of the particles is possible, therefore also no decay of space charge effects exists. 
The expansion continues until infinity if it is not stopped. 
Even though we introduced an effective acceleration time $t_\tx{acc}$, this problem persists. 
Since we keep both setup and method constant, the relative behavior of the cut-off energies can still be taken into account, but the absolute value is overestimated. 
This overestimation is predictable and when applied makes the models directly comparable.
Lécz et al.~\cite{Lecz2013} have shown that the acceleration time $t_\tx{acc}$ cuts off the spectrum, such that it is a good approximation of 2D simulations.
The simulations have been verified with experiments as well, which have shown that the bias can be mitigated.
Furthermore, Sinigardi et al. \cite{Sinigardi2018} have shown further scalings between 2D and 3D cutoff energies.
Taking both arguments into account we can deduce, that there is a constant scaling factor from 1D to real-world experiments and also to 3D simulations.
Similarly, because of the lack of transversal particle movement, we cannot evaluate divergence opening angles in a 1D simulation.

\subsubsection{Data Discussion by Example}\label{sec:examplediscussion}
\begin{figure}
    \includegraphics[width=\linewidth]{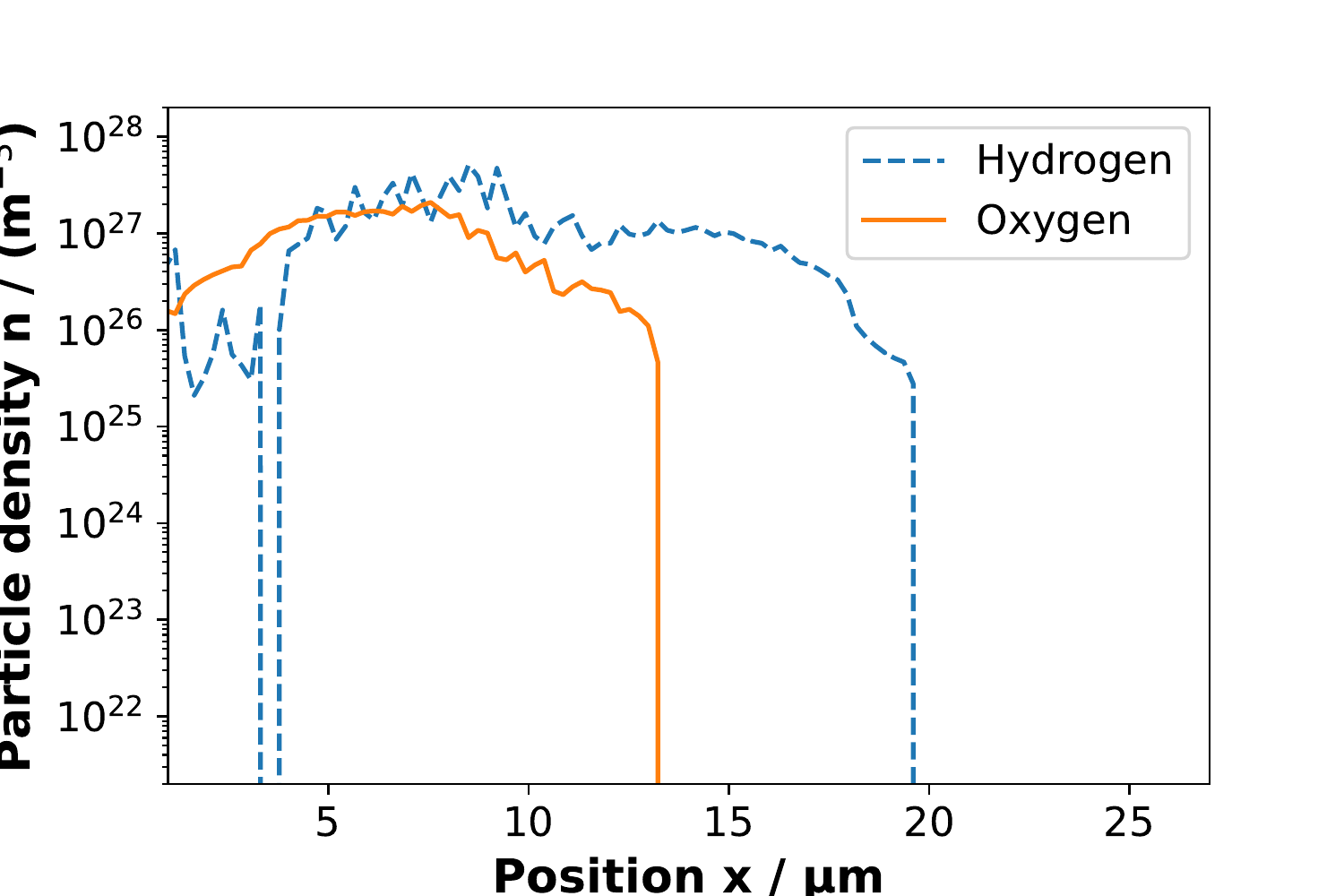}
    \caption{\label{fig:taccWater} Example PIC simulation of water leaf target TNSA experiment. The plot shows the particle distribution at the previously proposed acceleration time $t_\tx{acc}$ (Eq.~\eqref{eq:t_acc}).}
\end{figure}

We display an example of %one of the simulations in \autoref{fig:taccWater} and \autoref{fig:waterexamplee}.
the spatial distribution from the simulations in \autoref{fig:taccWater}.
An example of the energy spectrum is already displayed in \autoref{fig:picmulti-species}.

Firstly, in this simulation, we assumed that the target consists of regular water and is fully ionized by the implied laser pre-pulse.
Thus, the three species (e$^-$, H$^+$, and O$^{8+}$) are initialized with a density ratio of $10:2:1$ respectively such that overall neutrality is conserved. 
We display the species' positions at $t=t_\tx{acc}$ in \autoref{fig:taccWater}.
In this simulation the laser incidence angle is \SI{0}{\degree}, the target thickness is \SI{2}{\micro \meter} and the dimensionless laser amplitude is $a_0 = 20$. 
We observe the two ion species, H$^+$ and O$^{8+}$ at $t=t_\tx{acc}$. 
The figures show that the species have different positions at the measured time, which means that the species are accelerated separately by the sheath field.

The ion front position at the acceleration time for different species varies due to the different charge and mass values as mentioned in \autoref{sec:multi-species}.
Calculating the expected variation, following the relation from Huebl et al., for only fully ionized oxygen and hydrogen present, yields a scaling factor of $x_F^{O^{8+}} / x_F^{H} \approx 0.68$. 
The corresponding factor from \autoref{fig:taccWater} is $x_{F,\text{sim}}^{O^{8+}} / x_{F,\text{sim}}^{H} \approx \num{0.67\pm 0.01}$, where the uncertainty results from the binning. 

We can see that the general TNSA mechanism is still applicable.
Although the dynamics of the different particle species with each other are more complex, as we will see later, the general behavior appears to follow classical TNSA theory.
This is supported by the kinetic energy spectra of the ion species after acceleration, an example of which is shown in \autoref{fig:picmulti-species}.
The figure shows the energy spectra of hydrogen and oxygen ions, along with Mora's predicted ideal curve.

\subsection{Deep Learning Application}
We have to correlate the different simulations to each other and find relations and interpolations to allow for an optimization of the full setup.
We decided to use a neural network approach with fully connected feedforward topologies and built them in Keras~\cite{chollet2015keras} running inside of Tensorflow 2~\cite{tensorflow2015-whitepaper}.
For hyperparameter tuning, we used the Keras Tuner module~\cite{omalley2019kerastuner}.

\subsubsection{Model Training}
To predict a particle spectrum, two models are needed. The \textit{spectrum} model continuously maps $\{\tx{[physical parameters]}\}=\{E,\tx{mix}, E_\tx{L}, r_\tx{L}, \tau_\tx{L}, \tx{s/p-pol.}, \theta_\tx{L}, \lambda_\tx{L}, d_\tx{T}\}$ onto $\ln \left(\frac{\total n}{\total E} (E) \right)$ while a second \textit{cutoff} model only predicts the maximum energy (i.e. when to cut off the continuous spectrum from the first model).
We trained a reduced model pair, not taking deuterons into account for regular \ce{H2O}, and a full model pair containing different ratios between \ce{H2O} and \ce{D2O}.
The dedicated features of the PIC simulation can be seen better with the reduced model. 
We assume that this is a result of the lower number of input dimensions and therefore of the deviating degrees of generalization.

We essentially think of the energy spectrum as the graph of a continuous function $f$, the first model maps $\{ x, \tx{[system parameters]} \}$ to $f(x)$ while the second model predicts the point $x$ at which the graph gets cut off. 
Details about the training parameters and the procedure are given in Appendix \ref{sec:NetworkTraining}.
The reduced spectrum model has 6 hidden layers $(x\rightarrow 320 \rightarrow 288 \rightarrow 288 \rightarrow 256 \rightarrow 256 \rightarrow 320 \rightarrow 1)$ while the full spectrum model has 11 hidden layers with 460 neurons each. The cutoff models both have 8 hidden layers $(x \rightarrow 320 \rightarrow 284 \rightarrow 288 \rightarrow 512 \rightarrow 32 \rightarrow 480 \rightarrow 512 \rightarrow 32 \rightarrow 1)$. It is worth noting that the input dimension of the reduced spectrum model is one less than the full spectrum model since it does not include the mix parameter. All networks were fully-connected architectures with ReLU activations on their hidden layers.
We will now briefly discuss and evaluate the trained models:
\paragraph{Reduced Model Pair:}
The precision of the cutoff models, which attempt to map $\{\tx{[physical parameters]}\}$ onto $E_\tx{max}$ can be estimated rather easily.
For the reduced problem, the model achieved a mean squared error of \SI{8.93}{\mega\electronvolt\squared} on validation data (confer to appendix \ref{sec:NetworkTraining}), meaning the average error on the prediction of the hydrogen spectrum's maximum energy is projected to be around \SI{\pm2.99}{\mega\electronvolt}.

\begin{figure}
\includegraphics[width=\linewidth]{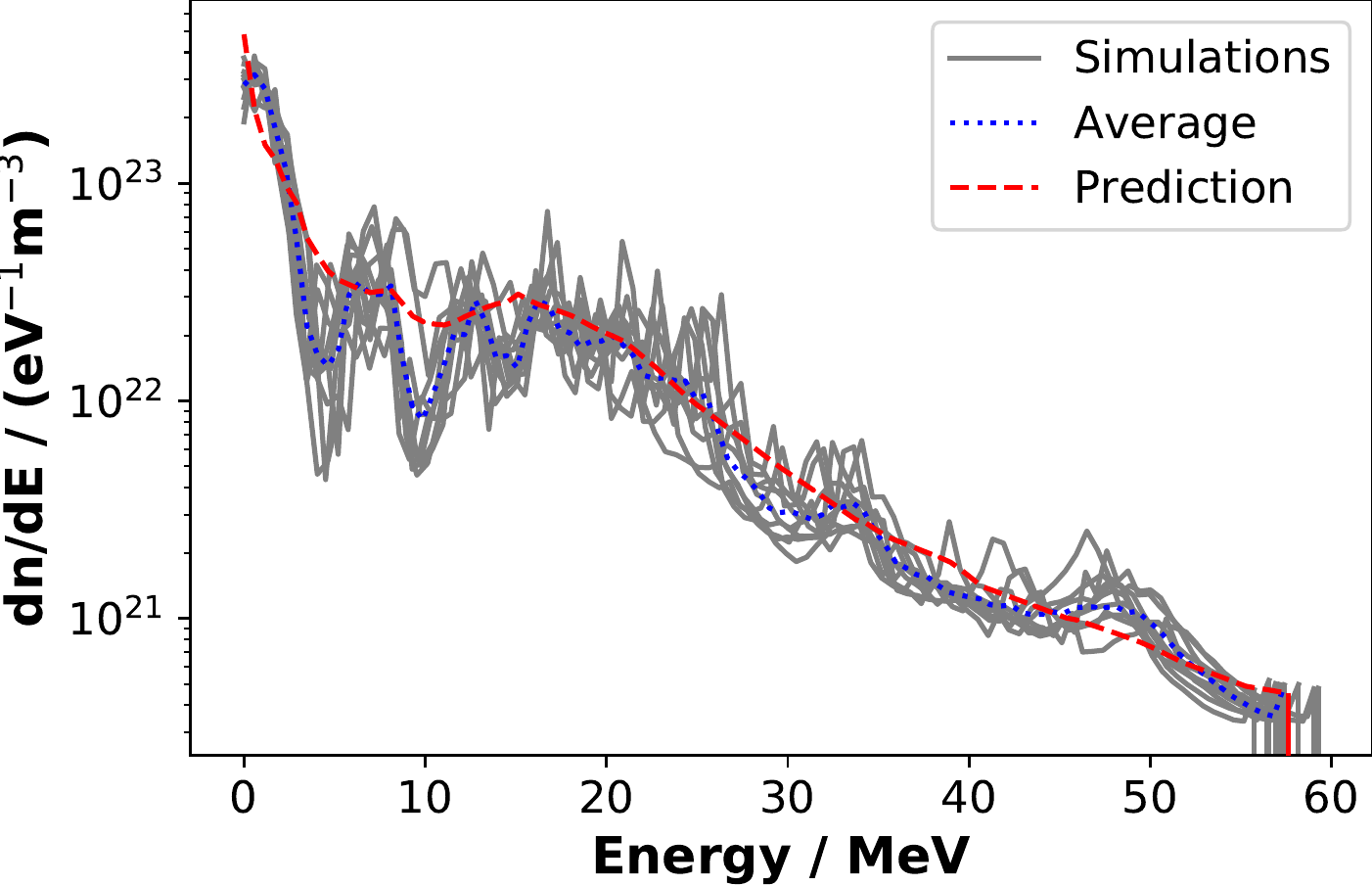}
\caption{\label{fig:SimVSModel}Reconstructed hydrogen ion energy spectra of ten simulations differing only in their random seed. 
The energy spectrum prediction by the trained neural network model is indicated with a red dashed line, while the average of the simulations is indicated by the blue dotted line. 
The curve is obtained from the reduced continuous model and is cut off at the maximum energy determined by the maximum energy model.}
\end{figure}

To more intuitively evaluate the reduced spectrum model's predicting capabilities and potential shortcomings, ten simulations with equal parameters (except for random seed) were computed such that their hydrogen ion energy spectra could be compared to the predicted spectrum of the model. A plot of all the spectra is shown in \autoref{fig:SimVSModel}.
The overall agreement of the model with the simulations is evident. The maximum energy predicted by the cutoff model falls centrally between the maximum energies of the ten simulations, only differing from the simulation average by \SI{0.2}{\mega\electronvolt}. Looking at more intricate features of the simulation spectra, however, it is clear that the model possibly generalized slightly too much. At around \SI{10}{\mega\electronvolt} a dip, possibly due to multi-species effects, can be observed in most of the simulations and yet is barely present in the model prediction. Generally, the fluctuations in the simulation spectra are greatly reduced in the neural network predicted spectrum. A reason for this is likely the sheer vastness of differing spectra the model was trained on. Since the parameter space for the training simulations was so large, the model had to generalize to many very different output spectra. 
\paragraph{Full Model Pair:}
The full model pair were trained exactly the same as the reduced model pair but with an additional parameter and a larger dataset. The full cutoff model converged with a mean squared error of \SI{7.25}{\mega\electronvolt\squared} on validation data resulting in a prediction error of \SI{\pm 2.7}{\mega\electronvolt} for the maximum energy of the hydrogen spectrum (confer to appendix \ref{sec:NetworkTraining}).

Again, as given above, the sensitivity of the spectrum model is more complicated to estimate. 
\begin{figure}
    \centering
    \includegraphics[width=\linewidth]{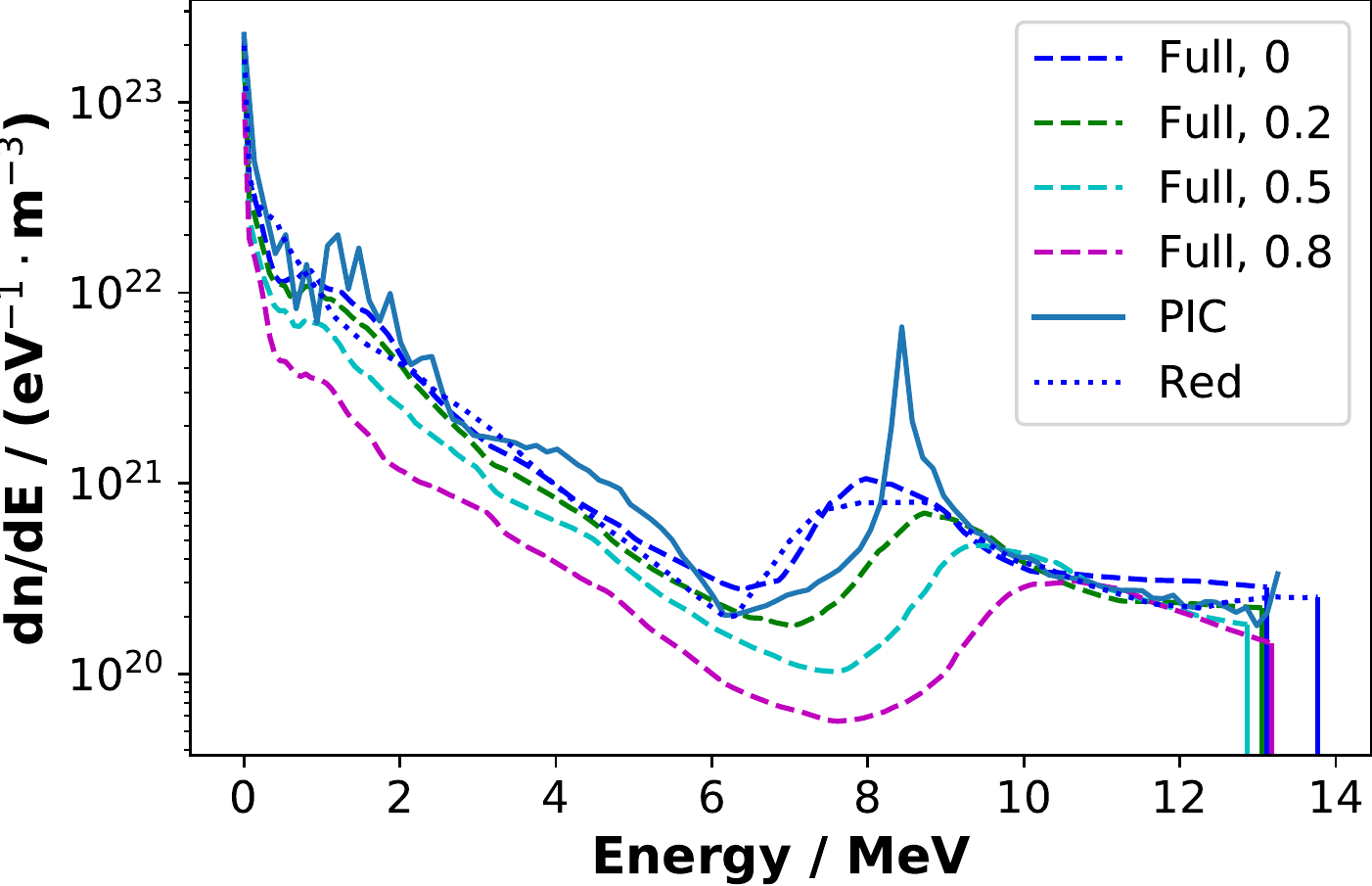}
    \caption{Model comparison for hydrogen spectra with reference PIC simulation. Dashed lines give the result for the full model, the number indicates the value for the mixture parameter. 
    The PIC reference (for mixture = 0) is displayed with a solid line and the reduced model with a dotted line.}
    \label{fig:compareNNspectral}
\end{figure}
In \autoref{fig:compareNNspectral}, we can see that both numerical models, the full and the reduced model do deviate from one another slightly, even if the mixture is set to zero. 
This is expected behavior since there is a statistical variation in the training of neural networks. 
Important to note is the deviation in the spectra for different mixture ratios. 
An influence of the mixture parameter on the spectrum is visible and it can be used to tune the spectrum.
The behavior for the full spectrum model is the same as the one for the reduced spectrum model presented in \autoref{fig:SimVSModel}, the model is generalizing to a specific degree and has an uncertainty of few MeV for the cut-off energy.

\subsubsection{Model Efficiency}
Calling the models in a Python code environment is similar to calling any other function and takes around \SI{20}{\milli\second} on a personal laptop.
This time is in stark contrast to the four hours on 16 CPUs taken to run a similar 1D PIC simulation on the HPC cluster.
To put this in perspective, we can run inference on the models roughly \num{720000} times, while one PIC simulation calculates.

We made other attempts at fitting the regression problem using various kernel combinations and Gaussian Process Regression~\cite{Bishop2006}, however, they never produced energy spectrum predictions that even came close to the neural network prediction seen in \autoref{fig:SimVSModel}, usually being off from simulations by orders of magnitude.
As expected, the adaptability of modern unparameterized machine learning methods such as neural network models stands out from other regressors. 

\section{Application of the Model\label{sec:applyDL}}
With a trained surrogate in hand, we were able to take advantage of the model to perform a numerical optimization of an experiment as well as evaluate our models' interpretability.

\subsection{Optimization of Parameters for Laser Plasma Interactions\label{sec:VegaOpti}}
In this section, we optimize a TNSA experiment with a water leaf target.
We aim at an ideal set of laser and target parameters and apply the previously obtained reduced machine learning model pair. 

We chose some base parameters that have a proven repetition rate of at least \SI{1}{\hertz}: Ti:Sa lasers with a central wavelength of \SI{800}{\nano\meter} and p-polarized laser light. 
Exemplary systems would be the VEGA-3 laser at the Centro De Laseres Pulsados in Salamanca, Spain (CLPU)~\cite{vega-laser} or DRACO laser at the Helmholt-Zentrum Dresden-Rossendorf~\cite{Schramm2017}.
Following the procedure in this section, the model can also be applied to any other system, if its parameters are inside the minimal and maximal physical parameters of our model (see \autoref{tab:physical_parameter_list}).
If the system's parameters are not included, our model could be expanded by retraining with additional data, using transfer learning~\cite{Spears2018}, or other modern domain adaptation methods~\cite{wang2018deep}.
The assumed initial parameters of the laser system are stated in \autoref{tab:vega_optimisation}. 

\begin{table}
    \caption{Table of the physical parameters to be optimized for the laser system.  
    Both initial and optimized values are shown. Rows in bold remained fixed during optimization. 
    The dimensionless laser amplitude $a_0$ also remained fixed during optimization to encourage the convergence towards non-trivial parameter combinations. $\eta_\tx{conv}$ is a measure for energy conversion efficiency (see Eq. \eqref{eq:e-conversion}), normalized to the initial parameter case.}
        \begin{tabular}{cccccc} \toprule
			No & Attribute & Initial & Optimized & Optimized  & Units \\
			 & & & ($E_\tx{max}$) & ($E \tx{-conversion}$) & \\ \midrule
			1   & Laser energy& 30 & 6.6 & 1.4 & \si{\joule} \\
			2  & Focus-FWHM  & 20 & 4.2 &  2   &\si{\micro\meter} \\
			3  & Pulse length & 30 & 149.9 & 137.6 &\si{\femto\second}\\
			\textbf{4}  & \textbf{Polarization}  & \textbf{p} &   \textbf{p}   &   \textbf{p}   &            \\
			5 & Incidence angle & 12.2 &  32.2 & 29.3 &\si{\degree}  \\
			\textbf{6}  & \textbf{Wavelength} & \textbf{800} & \textbf{800} & \textbf{800} & \textbf{nm}  \\
			7 & Thickness & 2 & 3.0 & 3.0 &\si{\micro\meter} \\\midrule
			& $E_\tx{max}$ & 13.8 & 51.5 & 51.2 & \si{\mega\electronvolt}\\ 
			& $\eta_\tx{conv}$& 1.0 & 7.8 & 41.3 & \\ \bottomrule 
		\end{tabular}
		\label{tab:vega_optimisation}
\end{table}

In this section, we investigate two different optimization goals.
The first goal is to find the maximum cut-off energy, while the second goal is to maximize the laser energy deposition into the plasma.

As mentioned we assumed polarization and central laser wavelength as fixed but otherwise allowed all parameters to change, as long as they stayed in the given physical constraints.
Since the obvious solution to maximizing output energy is to maximize input energy, the optimizations were computed under the constraint of a constant dimensionless laser amplitude $a_0$.
This ensures optimization by exploiting complicated relationships between the physical parameters of the system; a task that can only be feasibly solved with a rapidly callable model.

We implemented the optimization utilizing the \textit{SciPy} Python library~\cite{scipy} and the Byrd-Omojokun algorithm~\cite{lalee1998implementation} included in its \code{scipy.optimize.minimize} routine.
The Byrd-Omojokun algorithm allows us to include both, boundary conditions according to \autoref{tab:physical_parameter_list} as well as the equality constraint of constant $a_0$, to leverage the aforementioned non-trivialities of the system.

\begin{figure}
    \subfloat[Initial parameter selection.]{\includegraphics[width=.8\linewidth]{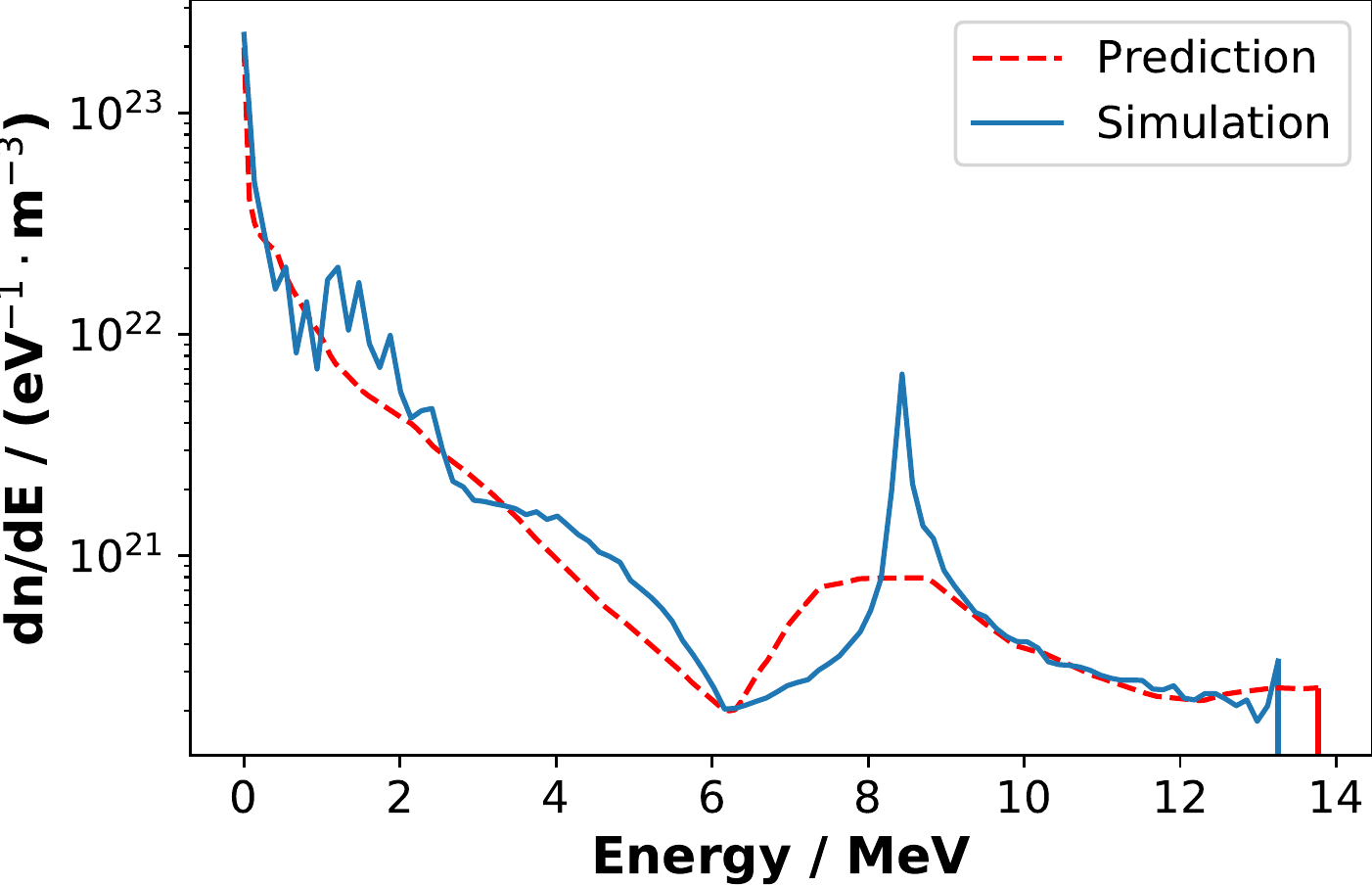}}\\
    \subfloat[Optimized parameter selection with respect to maximum ion energy.]{\includegraphics[width=.8\linewidth]{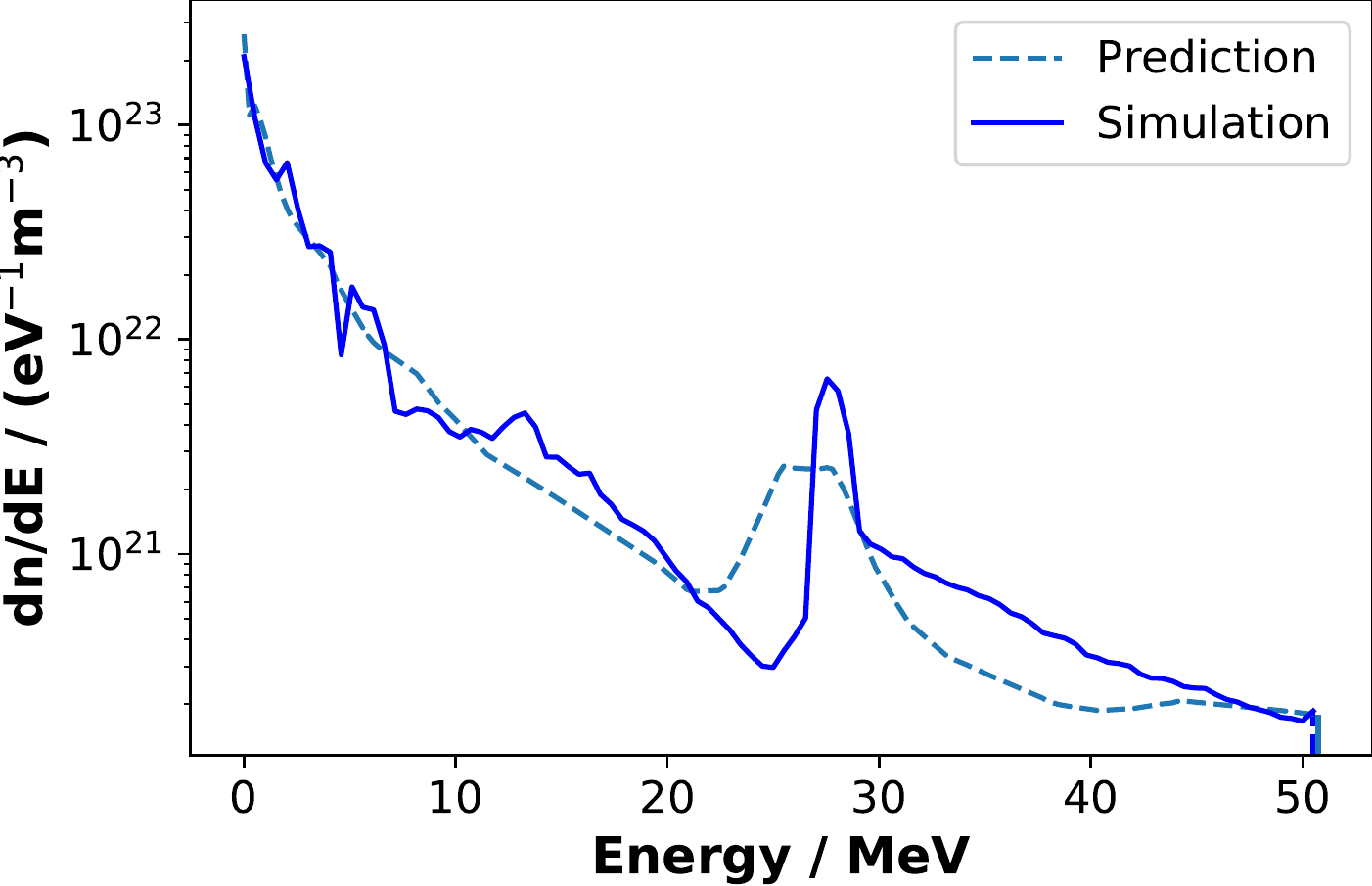}}
    \caption{\label{fig:VEGA-Optimised} Energy spectra of H-ions for a TNSA water leaf target experiment using both initial VEGA-3 as well as optimized parameters with respect to the maximum ion energy. Predicted spectra by the neural network model and spectra from a 1D PIC simulation are shown. The parameters are given in \autoref{tab:vega_optimisation}.}
\end{figure}

The optimized parameters are displayed in \autoref{tab:vega_optimisation}. 
The optimizer seems to have taken advantage of incidence angle-dependent absorption effects such as resonance absorption. 
Additionally, by dramatically increasing laser focus while simultaneously decreasing laser power (energy over time) the maximum ion energy could be optimized without changing the dimensionless laser amplitude $a_0$. 
Overall, the optimizer was able to increase the maximum ion output energy by a factor of roughly 4. 
The hydrogen energy spectra for these optimized parameters as well as for the initial parameters are depicted in \autoref{fig:VEGA-Optimised}.

A more intricate measure of a TNSA experimental system is the laser-ion energy conversion efficiency, i.e. the measure of how much of the laser's input energy gets transported into the accelerated particles. 
We thus consider the optimization of the ratio of the total kinetic energy of the ions $E_\tx{H}$ to the laser pulse energy $E_\tx{L}$:
% \begin{widetext}
\begin{align}
	\argmax_{x \in \{\tx{params}\}}\frac{E_\tx{H}(x)}{E_\tx{L}} &= \argmax_{x \in \{\tx{params}\}} \frac{1}{E_\tx{L}} \cdot \int_0^{E_\tx{max}} \frac{\total N}{\total E} \cdot E \, \total E \; , \label{eq:e-conversion}
\end{align}
% \end{widetext}
where $\frac{\total N}{\total E}(E,x)$ and $E_\tx{max}$ are given by the neural network models, and \{params\} is the set of all parameter combinations within the ranges specified in \autoref{tab:physical_parameter_list}.
It is important to note that the Smilei output gives $\frac{\total n}{\total E}$ which has to be scaled by a unit volume $V$ to arrive at the expression needed. 
For further explanation of how to arrive at the above integral term we refer to Appendix~\ref{sec:E-Appendix}. 
Here, we also allow the variation of laser energy $E_\tx{L}$, increasing the complexity of the problem.
The optimization described in Eq.~\eqref{eq:e-conversion} was carried out by solving the numerical integral using the composite trapezoidal rule and once again employing the Byrd-Omojokun algorithm. 

As seen in \autoref{tab:vega_optimisation}, despite having a slightly lower maximum ion energy than the first optimization task, the calculated energy conversion efficiency is more than five times greater. 
This gives a strong indication that laser coupling into the target in a laser-plasma experiment depends on the physical parameters of the system in a highly non-trivial way.

\subsection{Sensitivity Analysis\label{sec:sensitivity}}
Artificial neural networks are generally difficult to interpret, which is a drawback we have to accept. 
Nevertheless, the importance of specific parameters for a model can be evaluated. 
One way to quantify the impact of a model's input parameters on its output is to use variance-based global sensitivity analysis, also known as the Sobol' method. The corresponding sensitivity metrics are known as \textit{Sobol' indices}\cite{Sobol2001, Saltelli2002, Saltelli2010}. 
The Sobol' indices are calculated by Monte Carlo sampling of parameters and corresponding model outputs. This method is used to apportion the variance of the output to the inputs and their combinations.
The number of evaluations of our model is $N \times (2D + 2)$, where $D$ is the number of input features and $N$ is the number of samples drawn. 
$N$ is ideally selected as a power of 2, where we selected $2^{18} = \num{262144}$ drawn samples.

We used the PAWN method \cite{Pianosi2015} for a second sensitivity analysis to complement the Sobol' method due to its shortcomings for the higher order of the input features. 
The PAWN method uses a different approach for when the variance might not be a good measure for the outcome of a system. 
It utilizes the traits of the Cumulative Distribution Functions with similar Monte Carlo sampling as for the Sobol' indices, giving a different approach to determine the sensitivity of a model.
A combination of these two methods was also proposed by Baroni et al. \cite{Baroni2020}.

\subsubsection*{Reduced Model}
\begin{figure}
    \centering
    \subfloat[Total variation which explains the cut-off energy variation. ]{\includegraphics[width=.9\linewidth]{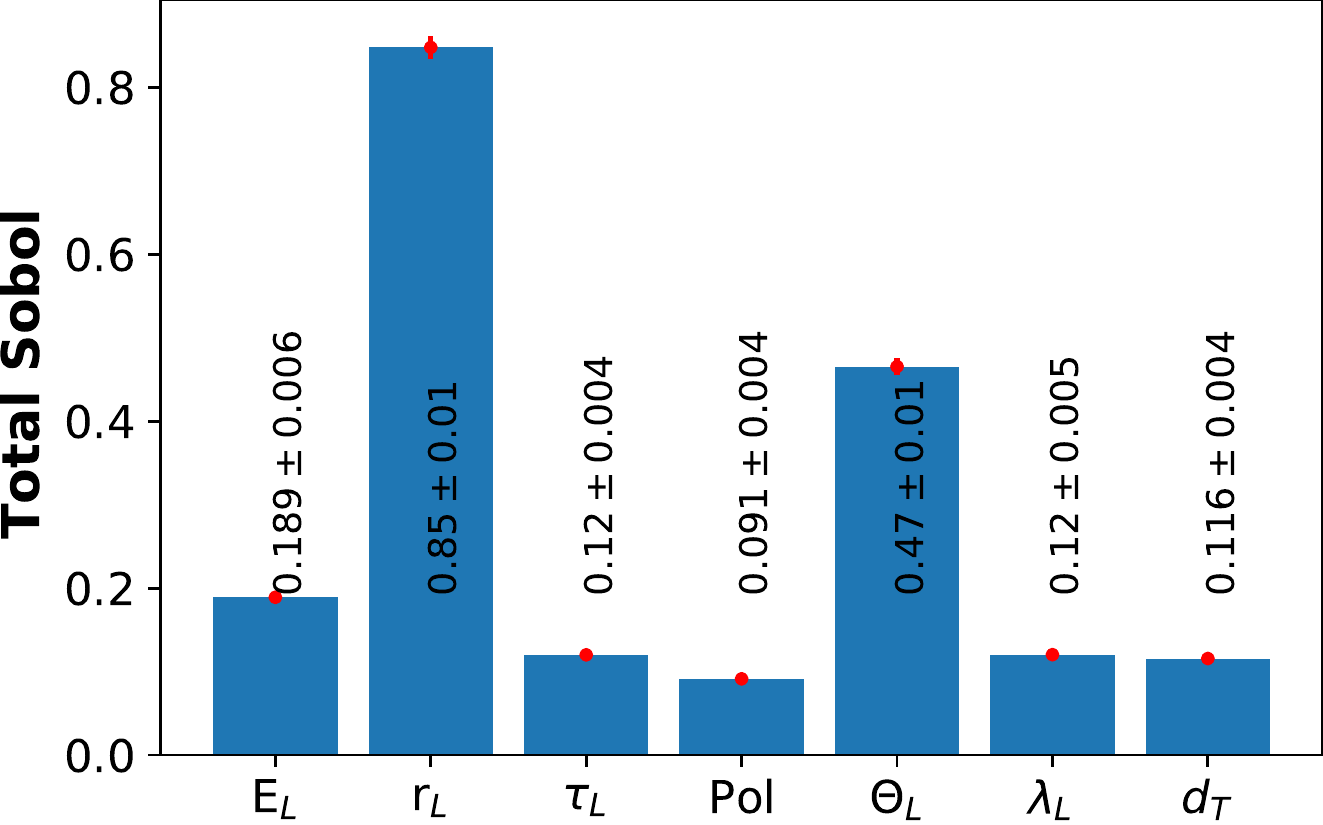}}\\
    \subfloat[Matrix of dependencies to explain the cut-off energy variation. The diagonal gives first-order Sobol' indices, while the lower gives the second-order Sobol' indices for the corresponding variables.
    The upper line is the numerical value, the lower line gives the corresponding error.]{\includegraphics[width=.9\linewidth]{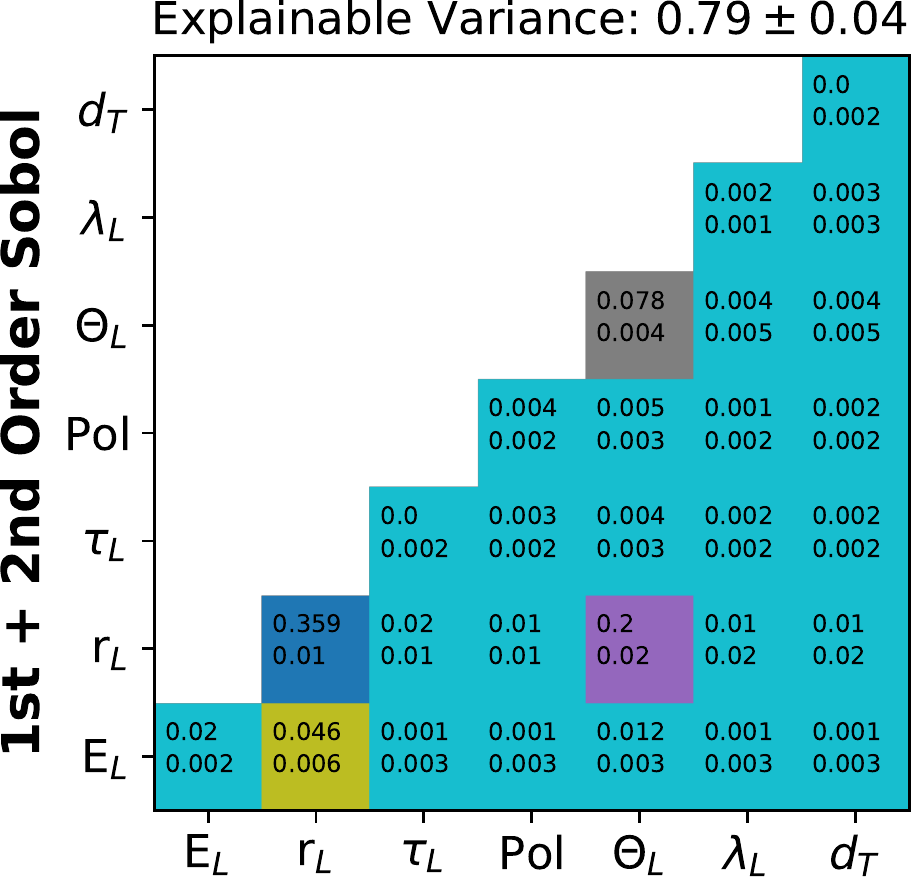}}\\
    \caption{Sobol' sensitivity analysis results showing the influence of various physical parameters on the cut-off energy of H-ions for a TNSA water leaf target experiment, for the reduced model utilizing only the \ce{H2O} data. Errors are given in the \SI{95}{\percent} confidence level.}
    \label{fig:ReducedSobolResult}
\end{figure}

The reduced cutoff model has 7 input features which are mapped to 1 output prediction for the maximal energy.
Our results of the Sobol' analysis for the reduced \ce{H2O}-only model are given in \autoref{fig:ReducedSobolResult}.
The larger the value of a Sobol' index is, the more influence the independent parameter has on the result.
The total Sobol' indices, normally referred to as $S_T$ give a measure of the total importance of the given features. 
The total Sobol' indices can neither describe how much of the variance is attributed to which combination of parameters nor are they normalized for the total expression. 
This is due to multiple counting of effects, e.g. if there is a second-order contribution for $\Theta_L$ and $r_L$, then this contribution is added to both of the values in the total representation. 
It doubles the counting for the second order, triples for the third order, and so on.  

Due to this complication, the determination of higher-order dependencies makes it necessary to display the first and second-order Sobol' indices, as done in \autoref{fig:ReducedSobolResult} (b). 
The values are displayed in a matrix, such that the interaction between $(x_i,y_j)$ can be displayed. 
The first order Sobol' indices are shown on the main diagonal ($x_i = y_j$).
It is evident from the plot, that the sum does not add up to 1, leaving approximately \SI{21}{\percent} of data variance unexplained. 
The consequence of this is, that even higher order interactions are necessary to fully explain the variation in our model.
Full calculation of higher orders has been omitted as it was deemed unfeasible due to the extreme computational cost for higher dimensions. 

\begin{figure}
    \centering
    \includegraphics[width=.9\linewidth]{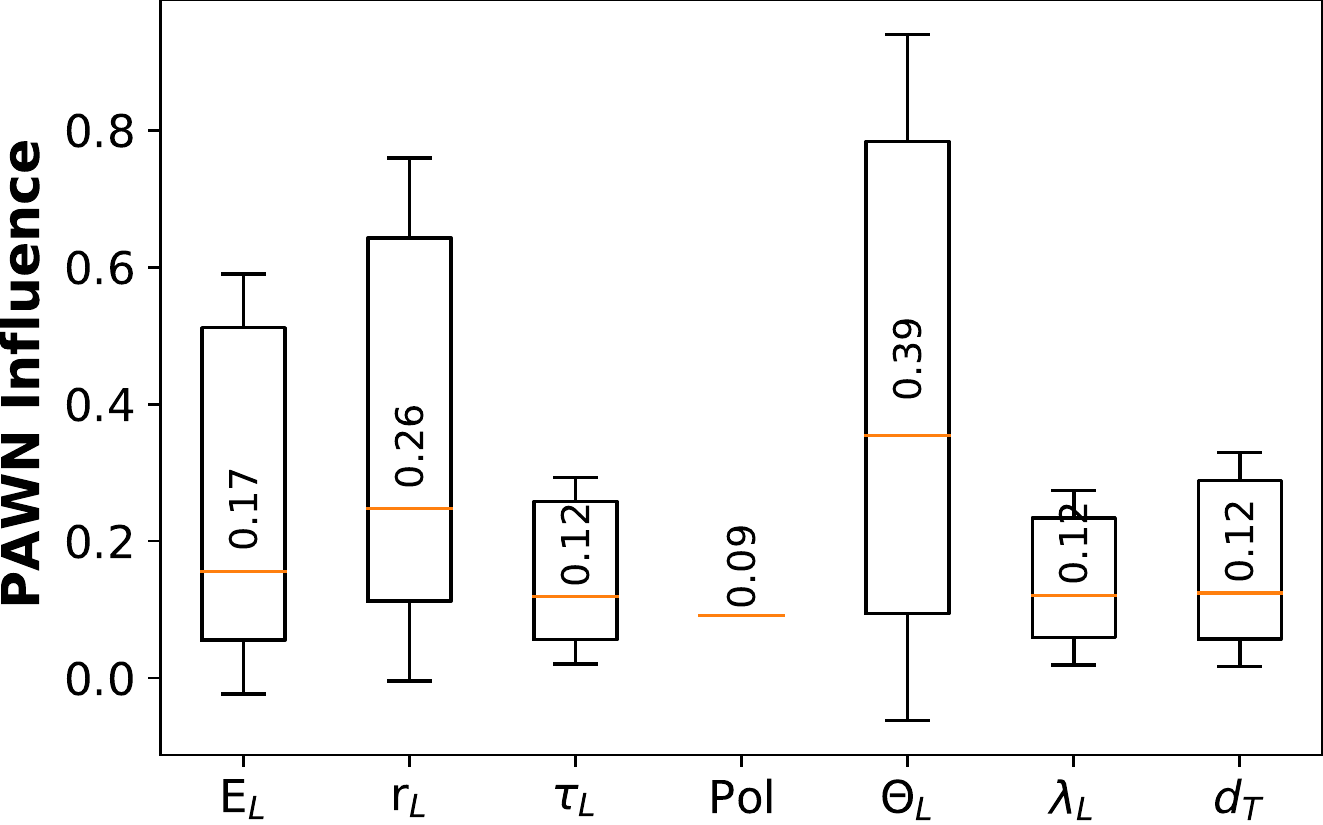}
    \caption{PAWN indices for the reduced model as a measure for parameter importance. Boxes consist of uncertainty value, minimum, median, maximum, and upper uncertainty value. The numerical value given is the median.}
    \label{fig:PAWN}
\end{figure}

The results of the PAWN method are displayed in \autoref{fig:PAWN}.
PAWN can only give us a measure of the full importance of the individual parameters. 
A subsequent division into main effects and higher order is not possible. 
\begin{table}
    \centering
    \caption{Importance ranking of the model parameters as calculated by the Sobol' and PAWN methods.\label{tab:OrderSensi}}
    \begin{tabular}{ccc}\toprule
    Importance & Sobol' & PAWN \\ \midrule
        1 & $r_L$ & $\Theta_L$\\
        2 & $\Theta_L$ & $r_L$\\
        3 & $E_L$ & $E_L$\\
        4 & $\tau_L$ & $\tau_L $ \\
        5 & $\lambda_L$ & $\lambda_L$\\
        6 & $d_T$ & $d_T$\\
        7 & Pol & Pol\\ \bottomrule
    \end{tabular}
\end{table}
The importance ranking from PAWN does not entirely match the order found by the Sobol' method but is rather close.
Both are listed in \autoref{tab:OrderSensi}.
If not the total, but the sum of first and second-order Sobol' is taken, then the first two features change places. 

The sensitivity analyses thus suggest that
higher-order interactions are important in this model and a simple optimization (e.g. maximizing only one quantity) is not sufficient. Our previously presented optimizations take this implicitly into account. 
Furthermore, the incidence angle and the irradiation area appear to be important.
The angle $\Theta_\tx{L}$'s high influence is expected, considering laser-ion absorption mechanisms, and is faithfully implemented into the 1D simulation space using the Lorentz boosted geometry (see Appendix \ref{sec:Lorentz}). 
While the third quantity, the laser energy, directly scales the laser's dimensionless amplitude $a_0$, the irradiation radius $r_\tx{L}$'s influence is more difficult to understand. 
The irradiation area is not directly represented in a 1.5D PIC simulation. However, since $a_0$ is dependent on $r_\tx{L}$ an indirect influence is included.

\subsubsection*{Full Model}
\begin{figure}
    \centering
    \subfloat[Total variation which explains the cut-off energy variation. ]{\includegraphics[width=.9\linewidth]{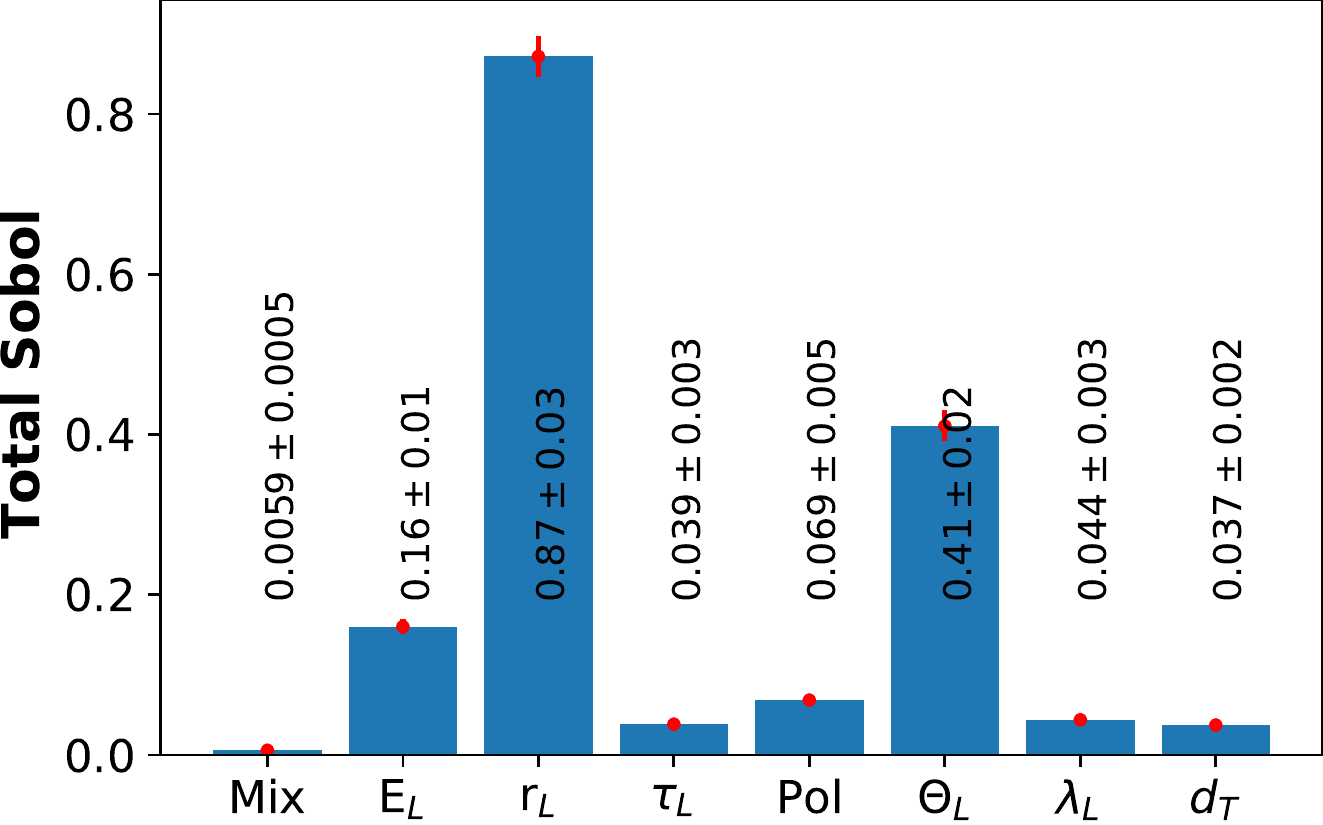}}\\
    \subfloat[Matrix of dependencies to explain the cut-off energy variation. The diagonal gives first-order Sobol' indices, while the lower gives the second-order Sobol' indices for the corresponding variables.
    The upper line is the numerical value, the lower line gives the corresponding error.   ]{\includegraphics[width=.9\linewidth]{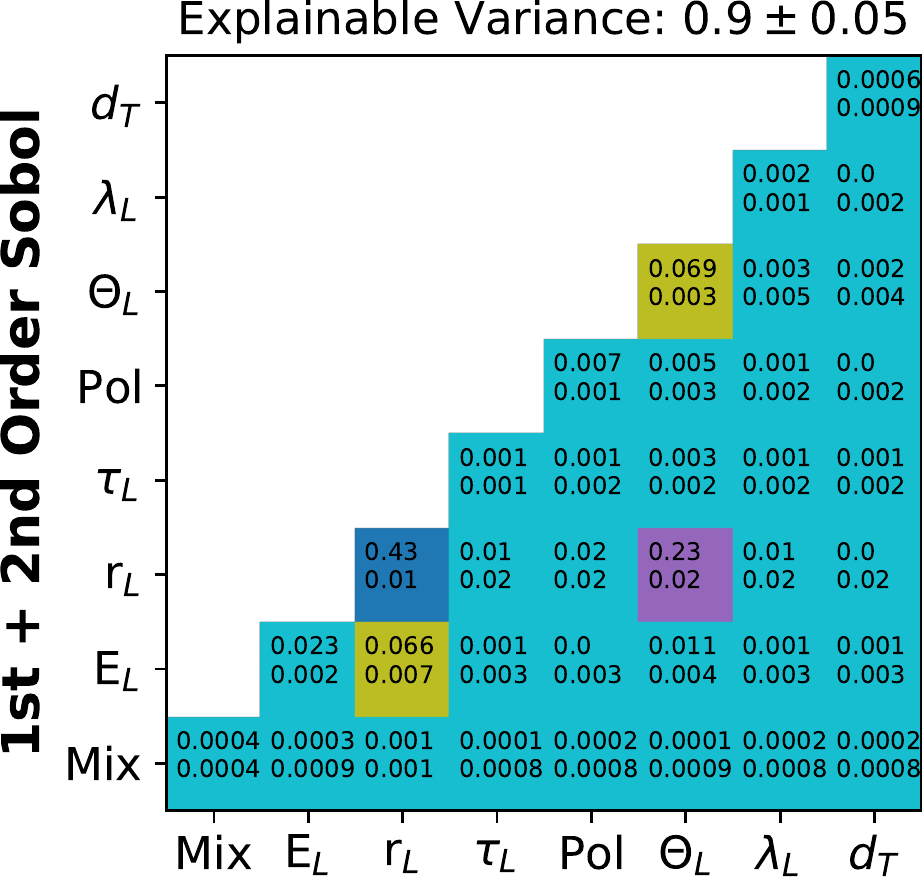}}\\
    \caption{Sobol' sensitivity analysis results showing the influence of various physical parameters on the cut-off energy of H-ions for a TNSA water leaf target experiment, for the full model utilizing only the \ce{H2O} data. Errors are given in the \SI{95}{\percent} confidence level.}
    \label{fig:FullSobolResult}
\end{figure}

\begin{figure}
    \centering
    \includegraphics[width=.9\linewidth]{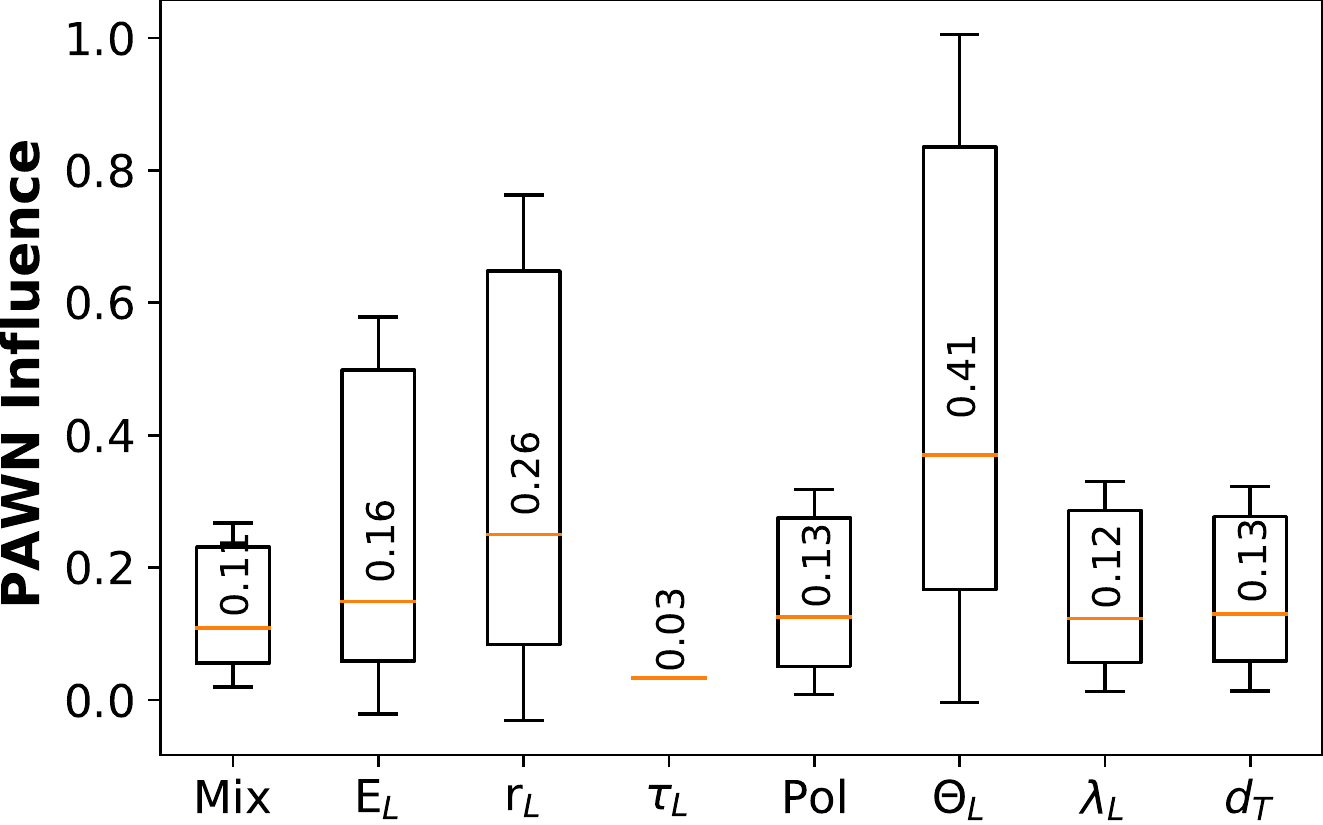}
    \caption{PAWN indices for the full model as a measure for parameter importance. Boxes consist of uncertainty value, minimum, median, maximum, and upper uncertainty value. The numerical value given is the median.}
    \label{fig:PAWNFull}
\end{figure}

Running the same analysis for the full cutoff model including the mixture parameters yields the results given in \autoref{fig:FullSobolResult} and \autoref{fig:PAWNFull}.
As can be seen in the display of the data, the mixture has, according to the Sobol' analysis, minimal if not zero influence on the maximum energy of the hydrogen component, while the PAWN analysis gives a higher influence. 
Furthermore, the variance of the output can be better explained in this model, than from the reduced model although the geometry did not change.

\subsubsection*{Sensitivity Analysis Discussion}
We performed a sensitivity analysis on our models and were able to evaluate the importance of the different parameters. 
We found evidence that the model describing the laser-plasma system is highly non-linear. It should be noted that a deep learning model approximates the physical system very well. It does not, however, provide a closed-form solution for the underlying physics, which would require further theoretical work.

The models apply regression for the simulated data and as such are able to reproduce a mean curve for the data, which is for example displayed in \autoref{fig:VEGA-Optimised} or in \autoref{fig:SimVSModel}. 
The spectrum models take the energy bin value into account and predict the continuum of accelerated ions. 
This means, that the dependency on the bin's energy would be included as well. 
An explainable analysis of taking all energy bin values into account then becomes infeasible as each bin would require a separate Sobol/PAWN analysis.

The cut-off energy of a TNSA spectrum is the main parameter investigated in the literature, which is for example analyzed by Zimmer et al. \cite{zimmer2021analysis}.
We have seen that we can get similar results to Zimmer et al. for the cut-off energy dependencies.
We see from the Sobol' analysis, that several parameters are of importance, therefore having only a single parameter to describe the cut-off is not sufficient. 
Since second-order Sobol' indices are not zero, we have to take them into account as well.

Since neither model is explaining the cut-off variation close to \SI{100}{\percent}, when only 1st and 2nd-order variations are taken into account, we can conclude that the calculated models require consideration of even higher order variations to describe an additional \SI{10}{\percent}-\SI{20}{\percent} of the cut-off variation.
Such a large reliance on higher-order interactions implies that simple scaling models are unideal for optimizations since these effects are not taken into account. A model capable of approximating highly non-linear effects, such as our neural network models, should thus be preferred. 

The Sobol' indices decompose the function into a unique space\cite{Sobol2001}, this could be used to construct a polynomial chaos expansion\cite{Ghanem1991} polynomial from it. 
This polynomial can describe the same amount of variation as indicated by first and second-order Sobol'. 
It is therefore neither a complete representation of our network-based models nor is it physically interpretable. 

\subsection{Interpretations}
Two major observations from the numerical study are of interest for the understanding of the modeled system.
        
The first observation is the deviation from the exponential Mora-like shape towards the plateau-like features as presented in \autoref{fig:picmulti-species}. 
An explanation of this effect is the particle-particle interaction inside the expanding plasma. 
The driver of this effect is the higher-mass particle species (heavy ions), which is accelerated later than the lower-mass particle species (protons). 
Heavy ions, with their higher inertia co-propagate with lower energy protons and interact via the Coulomb force. 
Due to the higher inertia, the protons are pushed away from the heavy ions, being accelerated as a result.
This effect is especially highlighted by 1D PIC simulations since no transverse particle movement is allowed. 
For higher order dimensions \cite{Huebl2020} or experimental data \cite{Alejo2014} the effect is less dominant and the transitions are smoother. 
If the particles are accelerated purely in the longitudinal direction, the divergence, which is given by the quotient of transversal and longitudinal momentum, can be reduced as well \cite[Eq. 2]{Schmitz2022}. 

The second observation is the high increase in energy absorption. 
The fraction of the energy passed onto the protons increases by a factor of about 42.
As shown in the validation for the angular Lorentz boost scheme in Appendix \ref{sec:Lorentz}, a large increase in the absorption efficiency is a result of the angle-dependent resonance absorption. 
Indications for this are displayed in \autoref{fig:absscan} and \autoref{fig:absVsCui}.
The optimization algorithm exploits this behavior directly and therefore finds ideal angle values. 
However, there are at least two sides to this coin. 
The goal of the approach was to describe the TNSA process in a model which allows for the optimization of the output depending on the input. 
Approaching this directly and analytically is not possible.
The time development of the governing Maxwell-Vlasov system, which already is a simplification using the collision-free case, can not be solved in closed form. 
No true relations for the cut-off energy, for example, have been derived so far.
To get as close to this ground truth as possible, and to become able to extract it at a later point (with sufficient experimental data), a complex numerical model must be used. 
In our case, we adopted an artificial neural network approach. 
Artificial neural networks have desirable properties as they have been shown to be universal function approximators \cite{leshno1993multilayer}. However, the explainability of such complex models has been a critical point in their analysis for some time.
The sensitivity analyses shown in \autoref{sec:sensitivity} were used as a way to mitigate the complexity and gain some explainability of the model.
The Sobol' indices method, or global variance-based method, underlines, that the interaction of the different parameters is of importance. 
First and second order can only explain 79\% of the models' variance (\autoref{fig:ReducedSobolResult} b). 
This means that higher-order dependencies of the input parameters are necessary to explain a significant part (21\%) of the variance.
The model can not explain which higher-order effect i.e. which combination of input quantities, is exactly responsible.
The model's goal is to allow engineering optimization of the TNSA process.
As a result of this optimization, this higher-order dependency was found.

\section{Conclusion}

In this study, we modeled and optimized a possible TNSA experiment using a liquid leaf target by employing a combination of Particle-In-Cell simulations and deep learning.
In agreement with previous studies \cite{Huebl2020, Alejo2014}, we have seen that the accelerated spectra from a multi-species target behave untypical in comparison to regular one-species TNSA which is described by Mora.

We developed surrogate models that replicate computationally costly PIC simulations using a deep learning approach. Deep learning is well-suited for optimizing complex systems. To take advantage of the trained models' inference speed, we used the Byrd-Omojokun algorithm to find an optimal parameter configuration for the system. This yielded a set of parameters that resulted in optimal maximum hydrogen energy (8 times greater than the initial parameters) and a set of parameters that resulted in optimal laser energy conversion efficiency (41 times greater than the initial parameters). We verified these findings with additional PIC simulations.

We applied sensitivity analysis methods to evaluate the influence of the different parameters and successfully identified the relevant ones.
We showed that such sensitivity analysis methods bear great potential for the understanding and quantification of physical dependencies when a closed-form solution is not known.

The data-based model that we developed can be extended in the future to improve predictions and better understand the system. This can be achieved by incorporating future experimental data for the liquid jet.
\section*{Author Declarations}
The authors have no conflicts of interest to disclose. 
\section*{Data availability statement}
Codes and data are available on request.
\section*{Funding Statement}
This work was funded by HMWK through the LOEWE center “Nuclear Photonics.”

This work is also supported by the Graduate School CE within the Centre for Computational Engineering at Technische Universität Darmstadt.

The results presented here are based on simulations, which were performed on the Virgo HPC cluster at the GSI Helmholtzzentrum für Schwerionenforschung, Darmstadt (Germany) in the frame of FAIR Phase-0.
\begin{acknowledgments}
We would like to thank the Smilei team for providing valuable discussions.
We would also like to thank Ion Gabriel Ion and Dimitrios Loukrezis for the helpful discussions on sensitivity analysis and the physical interpretation of artificial neural network models.
\end{acknowledgments}

\appendix

\section{Units and Dimensionality\label{sec:unitsAndDimensions}}
The simulations in this work were done using the particle-in-cell (PIC) method~\cite{Birdsall2004}. 
In the following section, we discuss the units and dimensions of the underlying Maxwell Vlasov system and extract a lower number of relevant parameters which give valuable physical insight. 
This is important to understand why 9 parameters were used in our model. 
The basis we construct in this chapter can be represented by the physical parameters we sampled for the simulation part of our study. 

\subsection{Basis Maxwell Vlasov System and Normalization\label{sec:basismaxwell}}

TNSA requires a high-intensity laser pulse to heat plasma electrons up to MeV temperatures. 
We assume that the mean free path is larger than the target thickness and the whole process can therefore be assumed as collision-free \cite{Mulser2010, Arber2015}.
If the process is collision-free, it can be characterized by the Maxwell-Vlasov system of partial differential equations.
\newcommand{\B}{\ensuremath{\vec{B}}}
\newcommand{\E}{\ensuremath{\vec{E}}}
\renewcommand{\j}{\ensuremath{\vec{j}}}
\begin{align}
    \nabla \cdot \B &= 0 \notag\\
    \nabla \cdot \E &= \frac{\varrho}{\varepsilon_0}\notag\\%4\pi\varrho\\
    \nabla \times\E &= - \frac{\partial \B}{\partial t}\label{eq:vlasov-eqs}\\
    \nabla \times\B &= \mu_0\j +\mu_0\varepsilon_0\frac{\partial\E}{\partial t}\notag\\
    0 &=\frac{\partial f_\alpha }{\partial t} + \vec{v}_\alpha\cdot\nabla f_\alpha + q_\alpha\left( \vec{E}+\vec{v}_\alpha\times\B \right) \cdot    \frac{\partial f_\alpha}{\partial \vec{p}}\notag
\end{align}
Solving these coupled equations efficiently with numerical methods makes it important to simplify relations. 
A normalization towards reference quantities is the first step:
\begin{align}
    t' &= \frac{t}{\tau} \quad\text{where }\tau\text{ is the pulse length} \\
    r' &= \frac{r}{L} \quad\text{where $L$ is the focus size on the target}\\
    p' &= \frac{p}{p_0} \quad\text{with}\quad p_0 = \frac{eE_0}{\omega_\tx{L}} \\
    \E &= E_0\hat{\E} \\ 
    \B &= \frac{E_0}{c} \hat{\B}
\end{align}
The charge distribution and the current can further be expressed by
\begin{align}
    \varrho &= \int \sum_\alpha f_\alpha \operatorname{d}^3\!\vec{p} \quad \text{and}\\
    \vec{j} &= \int \sum_\alpha f_\alpha \vec{v}_\alpha \operatorname{d}^3\!\vec{p} \quad\text{with}\quad \vec{v}_\alpha = \frac{\vec{p}}{m_\alpha}\left(1+\frac{p^2}{m_\alpha^2 c^2}\right)^{-1/2},
\end{align}where $f_\alpha$ denotes the charge density.
These can be normalized to a reference quantity as well by modifying $f_\alpha$ accordingly shifting all dimensions into the new $n_\alpha$
\begin{equation}
    \hat{f}_\alpha = \frac{f_\alpha}{n_\alpha}
\end{equation}

\subsection{Similitude Relations and Dimensional Reduction\label{sec:similitude}}
A system of equation can be simplified, by applying the Buckingham $\Pi$ theorem\cite{Buckingham1914}.
This theorem allows us to take the dimensional quantities of a problem into account and find underlying dimensionless quantities which reflect the actual physical meaning.

If the boundary and initial conditions are similar, then fewer dimensionless parameters than dimensional parameters can be found to fully represent this equation system.
This implies that the shape of the electromagnetic wave, defined by \B\, and \E, and the normalized charge density $\hat{f}_\alpha$, have to be similar.
Similar in this case means, that the governing function is the same except for some parameters which themselves can be derived using the Buckingham $\Pi$ Theorem as well. 

To ensure similarity in this work, a Gaussian profile was assumed for the electromagnetic wave, leaving the laser frequency $\omega_\text{L}$, the pulse length $\tau_\text{L}$, and the corresponding electric peak field $E_0$ as variable quantities.
The initial plasma distribution is defined as a homogeneous slab with particle density $n_0$ and a thickness $d_\text{T}$ with exponential decaying pre-plasma and skirt.
The exact relations are given in \autoref{sec:HPC}.

Keeping these initial conditions fixed allows us to apply the Buckingham $\Pi$ Theorem to the Maxwell-Vlasov system of equations. 
This results in dimensionless quantities $\Pi_i$ which are capable of describing all dimensional quantities inside the equation system. The dimensional quantities are given in \autoref{tab:pi_phys_list}.
\begin{table}
    \centering
    \caption{Overview of the dimensional quantities of the Maxwell-Vlasov EQS. Dimensions are listed in SI base dimensions. Buckingham $\Pi$ parameters are calculated by defining primary quantities which are used multiplicatively in each parameter.}
    \begin{tabular}{ccc}\toprule
        Quantity & Dimensions & Type\\\midrule
        $\tau$ & T$^1$ & \\
        $L$ & L$^1$ & \\
        ${q}/{m}$ & C$^1$ T$^1$ Mass$^{-1}$ & Primary\\
        $E_0$ & M$^{1}$ L$^{1}$ C$^{-1}$ T$^{-3}$ & \\
        $\omega$ & T$^{-1}$ & Primary\\
        $\mu_0$ & M$^{1}$ L$^{1}$ T$^{-2}$ C$^{-2}$ & Primary\\
        $\varepsilon_0$ & M$^{-1}$ L$^{-3}$ T$^{4}$ C$^{2}$ & Primary\\
        $n_\alpha$ & C$^{1}$ T$^{1}$ L$^{-3}$& \\\bottomrule
        \multicolumn{3}{l}{T: Time, L: Length, C: Current, M: Mass}
    \end{tabular}
    \label{tab:pi_phys_list}
\end{table}
Using these dimensional quantities, the following $\Pi_i$ are determined: 
\begin{align}
    \Pi_1 &= \omega_\tx{L}\tau_\tx{L} \\
	\Pi_2 &= \omega_\tx{L} L\sqrt{\mu_0\varepsilon_0}=\frac{\omega_\tx{L} L}{c}\\
	\Pi_3 &= \frac{q_\alpha n_\alpha}{\varepsilon_0 m_\alpha\omega_\tx{L}^2} = %
		\begin{cases}
			\dfrac{\hat{n}_\tx{e} e^2}{\varepsilon_0 m_\tx{e}\omega_\tx{L}^2}\quad \text{for electrons} \\[4ex]%\quad Q/4\pi  
			\dfrac{\hat{n}_\alpha Z_\alpha^2 e^2}{\varepsilon_0 m_\alpha \omega_\tx{L}^2}\quad \text{for ions}%\quad Q/4\pi 
		\end{cases} \\[2ex]
	\Pi_4 &= \frac{q_\alpha E_0}{m_\alpha\omega_\tx{L}}\sqrt{\mu_0\varepsilon_0} = %
		\begin{cases}
			\dfrac{E_0}{\omega_\tx{L} c}\dfrac{e}{m_\tx{e}} \quad \text{for electrons}\\[4ex]
			\dfrac{E_0}{\omega_\tx{L} c}\dfrac{Z_i e}{m_i} \quad \text{for ions}
		\end{cases}
\end{align}
The theorem also states that the number of resulting dimensionless parameters is lower than the number of dimensional parameters, reducing the complexity of the model.
\subsection{Interpretation of the dimensionless quantities}
These quantities are sufficient to describe and condition the EQS from a mathematical standpoint.
From a physical standpoint, this also creates valuable insight. 
$\Pi_1$ gives the number of $\E$-oscillations in the laser pulse and $\Pi_2$ the irradiation size of the laser.
$\Pi_3$ correlates the laser with the target since it is the ratio of the particle density to the critical plasma density defined by the laser. 
$\Pi_4$ describes the particle dynamic inside the laser's amplitude for each species. 
For electrons, $\Pi_4$ is identical to the dimensionless quiver velocity $a_0$.
The meaning is equivalent for the different ion species. 

Writing down the EQS and substituting the $\Pi$ parameters makes their importance apparent:
\begin{align}
    \tilde{\nabla} \cdot \hat{\vec{B}} &= 0 \\
    \tilde{\nabla} \cdot \hat{\vec{E}} &= \underbrace{\int\sum_\alpha \frac{\Pi_2\cdot\Pi_{3\alpha}}{\Pi_{4\alpha}} \operatorname{d}^3\tilde{\vec{p}}}_{= 0\text{ for } t = 0} \\
    \tilde{\nabla} \times \hat{\vec{E}} &= - \frac{\Pi_2}{\Pi_1}\frac{\partial \hat{\vec{B}}}{\partial\tilde{t}} \\
	\tilde{\nabla}\times\hat{\vec{B}} &= \frac{\Pi_2}{\Pi_1} \frac{\partial \hat{\vec{E}}}{\partial\tilde{t}} + \int\sum_\alpha\frac{4\pi\Pi_2 \Pi_{3\alpha}\tilde{\vec{p}}}{\sqrt{\tilde{p}^2\Pi_{4\alpha}^2 + 1}} \hat{f}_\alpha\operatorname{d}^3\tilde{\vec{p}} \\ 
	0 &= \frac{1}{\Pi_1}\frac{\partial\hat{f}_\alpha}{\partial \tilde{t}} + \frac{1}{\Pi_2}\frac{\Pi_{4\alpha}\tilde{\vec{p}}}{\sqrt{\tilde{p}^2 \Pi_{4\alpha}^2 +1}} \frac{\partial \hat{f}_\alpha}{\partial \tilde{\vec{r}}} \notag\\
	&+ \left({Z_\alpha}\hat{\vec{E}} + \frac{\Pi_{4\alpha}}{\sqrt{\tilde{p}^2 \Pi_{4\alpha}^2 + 1}} \tilde{\vec{p}}\times \hat{\vec{B}} \right)\frac{\partial\hat{f}_\alpha}{\partial \tilde{p}} 
\end{align}
If these parameters are constant, then the equations are all the same and therefore behave the same.
This results in the same time development of the system and therefore yields the same results. 
One can say that for constant $\Pi_i$ areas of iso-dynamics exist, finally simplifying any model approaches by reducing the dimensions to be examined. 
Models therefore only need these 4 parameters to precisely determine a system.

\subsection{Correlating dimensionless parameters and simulation input}
The system, therefore, has a dedicated number of $\Pi$ Parameters which have to be taken into account: 
$\Pi_1$ and $\Pi_2$ are laser-relevant quantities and therefore particle species independent.
$\Pi_3$ and $\Pi_4$ describe quantities of the particle species, therefore introducing a multiplicity in the parameter, denoted by $\alpha$.
In the case investigated here, the multiplicity is 4: electrons, oxygen, hydrogen, and deuterium.
The system is initialized with the same spatial distribution function $\hat{f}_\alpha$ for each species. 
To mimic ionization and mixture some conditions apply: 
\begin{equation}
    N_O = 2\times N_{H/D} \quad\text{and}\quad N_e = Z^{\text{eff}}_O + Z_{H/D}
\end{equation}
Taking these assumptions into account resolves the multiplicity and the corresponding $\Pi$s can be expressed with a multiplicative factor.
The construction, including the multiplicative factors and the needed parameters, are given in \autoref{tab:pi_param_const}.
Red and green mark the relevant, varying parameters to be taken into account. 

\begin{table}
    \centering
    \caption{Construction of $\Pi$ parameters.}
    \begin{tabular}{lll}\toprule
        Parameter & Definition \\ \midrule
        $\Pi_1$ & $\textcolor{black}{\omega_L} \tau_\tx{L}$  \\
        $\Pi_2$ & $\textcolor{black}{\omega_L} L /c$  \\ \midrule
        $\Pi_{3O}$ & $\textcolor{red}{Z}^{\text{eff}}_O /m_O \times n_O/\textcolor{black}{\omega_L}^2$  \\
        $\Pi_{3e}$ & $\Pi_{3O} \times \left(\textcolor{red}{Z}^{\text{eff}}_O + 2\right) \times \frac{m_O}{m_e} \cdot \frac{q_e}{e\cdot\textcolor{red}{Z}^{\text{eff}}_O} $  \\
        $\Pi_{3H}$ & $\Pi_{3O} \times 2\text{\textcolor{green}{mix}} \times \frac{m_O}{m_H} \cdot \frac{q_H}{e\cdot\textcolor{red}{Z}^{\text{eff}}_O}$  \\
        $\Pi_{3D}$ & $\Pi_{3O} \times 2\left(1-\text{\textcolor{green}{mix}}\right) \times \frac{m_O}{m_D} \cdot \frac{q_D}{e\cdot\textcolor{red}{Z}^{\text{eff}}_O} $\\\midrule
        $\Pi_{4e}$ & $ q_e/m_e \times E_0 /\textcolor{black}{\omega_L} c $ ($a_0$)\\
        $\Pi_{4H}$ & $\Pi_{4e} \times \frac{m_e}{m_H} \cdot \frac{q_H}{q_e} $ \\
        $\Pi_{4D}$ & $\Pi_{4e} \times \frac{m_e}{m_D} \cdot \frac{q_D}{q_e}$ \\
        $\Pi_{4O}$ & $\Pi_{4e} \times \frac{m_e}{m_O} \cdot \frac{\textcolor{red}{Z}^{\text{eff}}_O}{q_e}$ \\
        \bottomrule
    \end{tabular}
    \label{tab:pi_param_const}
\end{table}
\subsubsection*{Mapping physical to dimensionless parameters}
As stated in the main body of this work, several physical input quantities are used. 
They are chosen based on keeping datasets consistent and comparable. 
Therefore some parameters are sampled which do not exist in 1D. 
This also ensures that the data can be broken down into the $\Pi_i$ with the following relations. 
It is important to note that from the $\Pi_4$ possibilities only the electron variant (equivalent to $a_0$) has to be passed to the PIC code. 
\begin{align}
    \Pi_1 &= \frac{2\pi c \tau_\text{L}}{\lambda_\text{L}} \label{eq:map1}\\ 
    \Pi_2 &= \frac{2\pi \text{FWHM}}{\lambda_\text{L}} \label{eq:map2}\\
    \Pi_3 &= \frac{Z_\text{eff}^2 e^2 n_O \lambda_\text{L}^2}{\varepsilon_0 4\pi^2 m_O c^2} \label{eq:map3}\\
    \Pi_4 &=  \left(\frac{E_\text{L}\lambda_\text{L}^2\sin\left(\Theta_\text{L}\right)}{\tau_\text{L}\pi \text{FWHM}^2 \cdot 1.37\times 10^{18}} \right)^2 \label{eq:map4}
\end{align}

This culminates in a needed dimensionality of 9 for the list of parameters:
4 $\Pi_i$, 2 parameters to deal with the ambiguity $Z^\text{eff}_O$ and the mixture parameters, 1 parameter for the plasma slab $d_T$ (particle density is fixed) and 2 for dealing with laser's polarization: Selection whether p / s linear polarization and to make a difference, variation of the incidence angle $\Theta_L$.

Taking the mapping into account (Eqs. \eqref{eq:map1}--\eqref{eq:map4}). 
A proper physical sampling includes:
\begin{compactenum}
    \item Ionization of oxygen
    \item Mixtures (deuterium vs hydrogen)
    \item Laser Polarization
    \item Laser Energy / Joule
    \item Laser Pulse Time / second
    \item Laser Irradiation Size / micron
    \item Laser Wavelength / meter
    \item Laser Incidence Angle / Degree (to the plasma normal)
    \item Plasma Slab Thickness
\end{compactenum}

These 9 parameters have the same dimensionality as the parameter space calculated by the $\Pi_i$ which is necessary since the construction of dedicated quantities (especially the electric field of the laser) cannot be determined easily and a composition of these parameters has to be taken into account. 

\section{Parameter Ranges \label{sec:paramRanges}}
    As mentioned in \autoref{sec:laserpulseparams} two paradigms are relevant for the selection of the parameter ranges. 
    
    The class of Petawatt laser systems we based our work on uses mainly linearly polarized laser light and is capable of varying the incidence angle. 
    Taking this into account we get two possibilities for the laser's polarization: s-polarization and p-polarization. Similarly, we can get angles from \SI{0}{\degree} to less than \SI{90}{\degree}. 
    At \SI{90}{\degree} the laser is not hitting the target and is traveling parallel to the plasma surface, we, therefore, chose to cut the interval at \SI{85}{\degree}.
    
    The laser energy was selected to cover a large area to increase the comparability of the model with different laser systems. 
    At the conceptualization phase of this study, it was unreasonable to assume high repetition rate experimentation with much larger systems (e.g. GSI's PHELIX system\cite{Bagnoud2009, Busold2014}), since the currently achievable repetition rates were too low.
    This might change in the future, such that higher energies are realistic, and the model base has to be expanded under such cases. 
    
    The most critical parameter is the pulse length. As our reference value, we selected the FWHM in the time domain of a pure Gaussian pulse. 
    Firstly, the approximation of a pure Gaussian pulse is not necessarily true for a technically implemented laser. 
    If the FWHM according to \autoref{eq:timeEnvelop} is not used, then the value has to be adjusted accordingly.
    Due to calculation time issues of the underlying PIC models, we reduced the selected times to the interval from \SI{15}{\femto\second} to \SI{150}{\femto\second}. 
    We know that this time can be significantly larger, but high repetition systems can operate with low pulse length variables. 
    We also acknowledge, that our lower simulation border for the time is close to the bandwidth-limited pulse limit, but we wanted to have some lower data points to force the interpolation into good behavior and therefore mathematically overshot into the lower regime.
    The upper pulse length boundary is also the first parameter we want to increase in further studies since technical laser systems do need a larger pulse length to apply this model. 
    
    The focus FWHM was then sampled according to the definition of the laser $a_0$ conditions which we applied to our parameters. 
    We made sure to stay in the TNSA regime and prevented 
    the laser $a_0$ to be smaller than one and keep the focus still realistically small with \SI{2}{\micro\meter} as the lower range cut-off. 
    
    The selected wavelengths are larger than those used in engineered systems and also are somehow continuously sampled from this larger range. 
    The reason for this is the importance of the wavelength parameter following the similitude relations, which are dependent on the laser wavelength in every component.
    
    Concerning the target, we chose the thickness according to the parameters of the physical implementation of a liquid jet, which is currently under development.
    The mixture can only vary between 0 and \SI{100}{\percent}.
    Again, the effective charge of the particles plays an important role.
    We started with fully ionized oxygen and found traits of the multi-species effect during our investigation. 
    While discussing the results, we generalized the effective charge discussion but were not able to properly simulate different effective ionization levels.
    This is due to a lack of proper ionization models (also beyond the scope of this study) and limited numerical resources.
    This would also be a parameter that could be further improved in additional studies.

\section{Transversal Lorentz Boosted 1.5D PIC Simulations \label{sec:Lorentz}}
Modeling oblique laser incidence onto a target is inherently at least a 2D problem, which requires substantially more computational power than a similar 1D geometry to simulate. Bourdier~\cite{Bourdier1983} thus proposed a method in which a relativistic Lorentz boost is applied to the frame of reference in the simulation. This method has later been employed by Gibbon et al.~\cite{Gibbon1999} in a PIC code. 

\begin{figure}[h]
    \includegraphics[width=\linewidth]{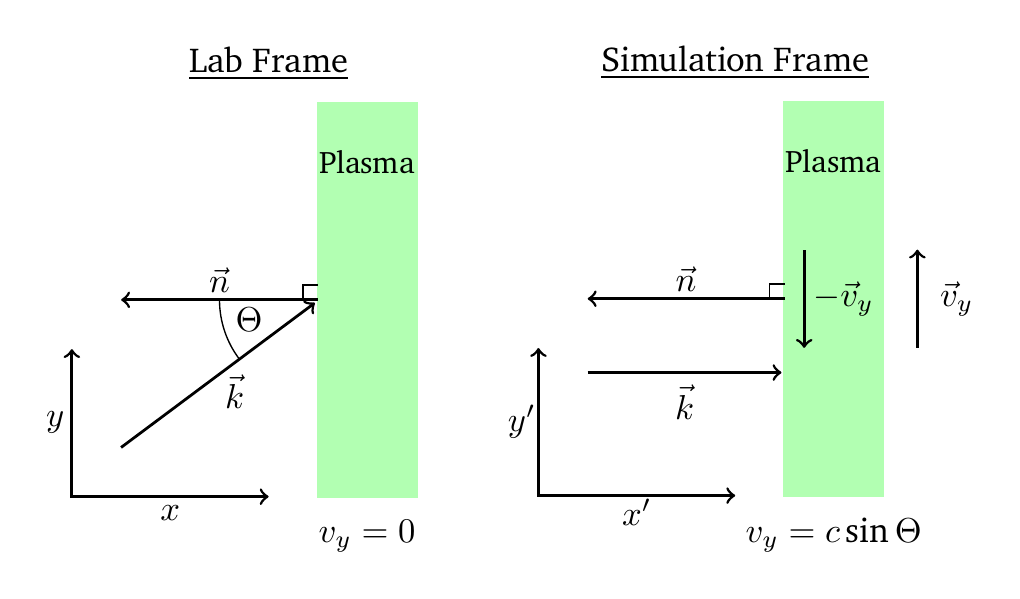}
    \caption{\label{fig:lorentzscheme} Schematic of the Lorentz boosted simulation frame versus the implied lab frame. In the simulation frame, the laser appears to be at normal incidence onto the target while the particles appear to drift in negative $y$-direction with velocity $v_y$.}
\end{figure}

Here, we would like to present the implementation of this technique yet again for a modern PIC code while also correcting some mistakes in the calculations by Gibbon et al.. 
A schematic of the general principle is shown in \autoref{fig:lorentzscheme}. 
To obtain the results in the lab frame a back transformation must be applied to the diagnostics obtained from the simulation.
For finding the transformations, a simple Lorentz boost in $y$-direction by the velocity $v_y = c \cdot \sin(\theta)$ is applied. In matrix form, this can be represented by:
	\begin{align}
		\Lambda &= \begin{pmatrix} 1/\cos(\theta) & 0 & -\tan(\theta) & 0\\
			0 & 1 & 0 & 0\\
			-\tan(\theta) & 0 & 1/\cos(\theta) & 0\\
			0 & 0 & 0 & 1
		\end{pmatrix} \; ,\label{eq:lorentzOne}
	\end{align}
	since $\gamma_0 = \left(1- (v_y/c)^2\right)^{-1/2} = 1/\cos(\theta)$. This transformation matrix can be used to transform all the quantities of the particles and the electromagnetic fields.
	Indicating quantities in the transformed system with a prime, we find after carrying out all transformations
	\begin{align}
		\begin{split}
			k'_y &= 0\\
			\omega_\tx{L}' &= \omega_\tx{L} / \gamma_0\\
			a_0' &= a_0 \; ,
		\end{split}
	\end{align}
	where $k'_y$ is the y-component of the wave vector in the boosted system, showing that, indeed, the laser is now at normal incidence. Note that the dimensionless laser amplitude $a_0$ is invariant under the transformation~\cite{Gibbon1999}. Further, denoting PIC code units with a tilde, we find
	\begin{align}
		\begin{split}
			\tilde{x}' &= \tilde{x}/\gamma_0\\
			\tilde{t}' &= \tilde{t}/ \gamma_0^2
		\end{split}
	\end{align}
	giving a re-scaling of both the simulation time as well as the cell grid. The initial particle density is also affected by
	\begin{align}
		\tilde{n}_0' &= \tilde{n}_0 \cdot \gamma_0^3 \; .
	\end{align}
	With these conditions, the particles can be initialized in the boosted frame. The relative velocity $v_y$ is added as a permanent drift which is handled and relativistically added to the particles by the code.\\
	From the diagnostics in the simulation, we can obtain desired quantities via a back transformation. For the particle kinetic energies, we find using the energy-momentum relation:
	\begin{align}
		\begin{split}
			\tilde{E} &= \gamma_0 (\tilde{E}' + \tilde{v}_y\tilde{p}_y')\\
			&= \gamma_0 \left(\frac{1}{m_\tx{e} c^2} \sqrt{{p'}^2 c^2 + (m_0 c^2)^2} + \tilde{v}_y\tilde{p}_y' \right)\\
			\Rightarrow E_\tx{kin} &= \gamma_0 m_\tx{e} c^2 \left( \sqrt{\tilde{p}'^2 + \frac{m_0^2}{m_\tx{e}^2}} + \tilde{v}_y \tilde{p}_y' \right) - m_0 c^2 \; ,
		\end{split} \label{eq:ekindirect}
	\end{align}
	where $m_0$ is the particle rest mass. Noting that $\tilde{B}_x' = 0$, we can find relations to recover the fields of the laser. The non-zero fields are:
	\begin{align}
		&\text{For s-polarization:}\nonumber \\
		\begin{split}
			&\tilde{E}_z = \tilde{E}_z'\\
			&\tilde{B}_x = \tilde{v}_y \tilde{E}_z'\\
			&\tilde{B}_y = \tilde{B}_y' / \gamma_0
		\end{split}\\
		&\text{For p-polarization:}\nonumber \\
		\begin{split}
			&\tilde{E}_x = \tilde{E}_x' - \tilde{v}_y \tilde{B}_z'\\
			&\tilde{E}_y = \tilde{E}_y' /\gamma_0\\
			&\tilde{B}_z = \tilde{B}_z' - \tilde{v}_y \tilde{E}_x'
		\end{split}
	\end{align}
	Using the field transformations and assuming that reflection at the plasma surface does not change polarization, we find for the absolute magnitude of the Poynting vector:
	\begin{align}
		|\vec{S}| &= |\vec{S}'| \cdot \gamma_0^2 \; ,
	\end{align}
	with which the relative absorption of the laser into the plasma can be calculated by dividing the incoming Poynting flux by the outgoing Poynting flux.
	
	It should be noted that while the Lorentz boosted frame method can replicate incidence angle-based behavior, it cannot replace a 2D or even 3D simulation on all accounts~\cite{Gibbon1999}. 
	Firstly, in the general case, all physical quantities depend separately on the transformed coordinates $x,y,z,t,p_x,p_y,p_z$. 
	Thus, the Lorentz-boosted simulation can only be used for a problem independent of $y$ and $z$. 
	Additionally, reducing the geometry after the boost to 1D limits the spatial dynamics of the particles. 
	Since only the $x$-axis is present, all particles (while having 3D velocities) can only move along a straight line (i.e. have only 1 spatial dimension). 
	This disregards the angular spread at the back of the target such that the particles can be accelerated for longer times and thus end up with higher energies compared to a similar 2D simulation. 
	Distinctly 2D effects such as hole boring can also not be modeled accurately.
	
\begin{figure}
        \includegraphics[width=\linewidth]{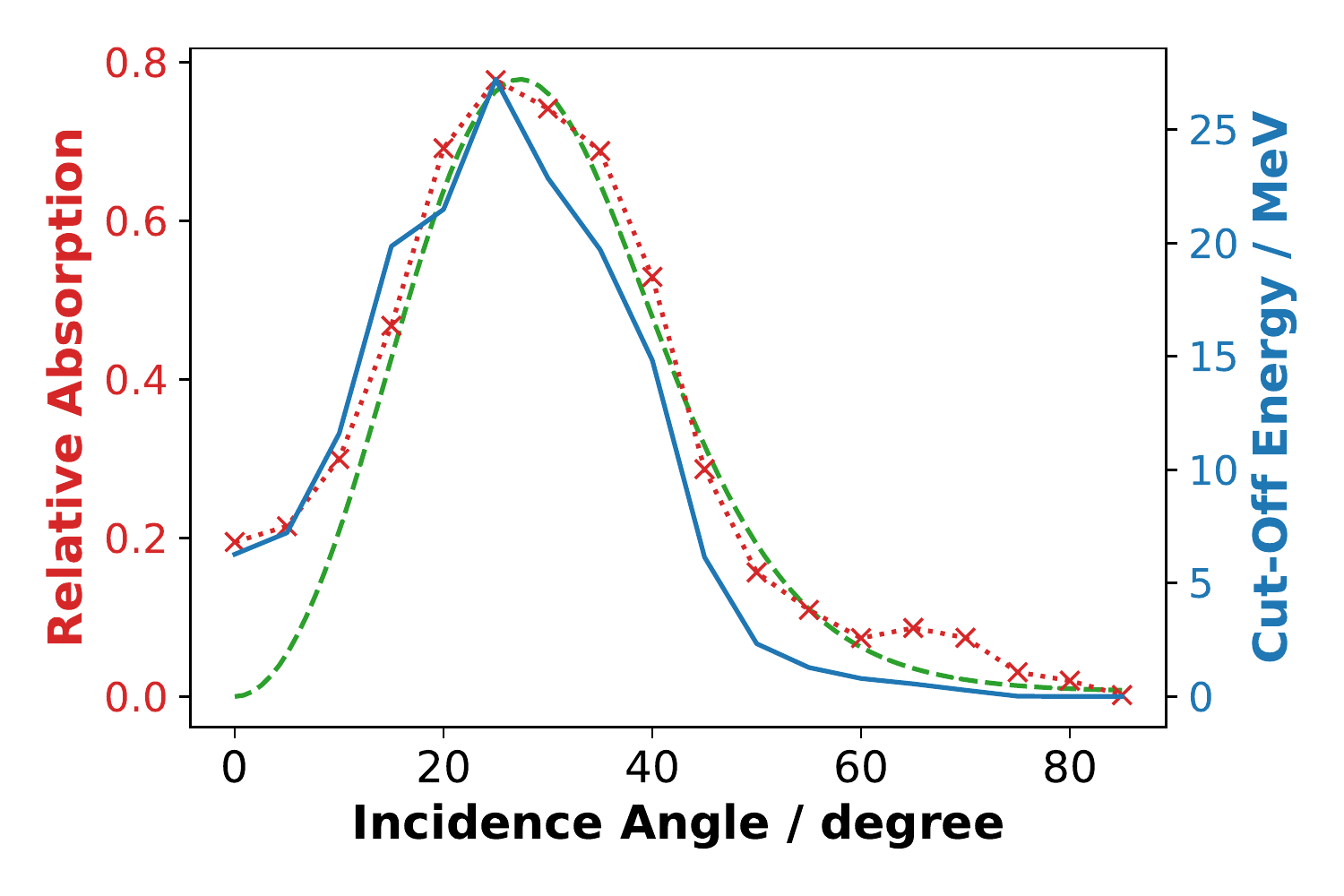}
        \caption{\label{fig:absscan} Simulation of laser incidence angles between \SI{0}{\degree} and \SI{85}{\degree}. The plot shows the incidence angle versus the relative absorption of a laser into a hydrogen plasma target (dotted red line)  as well as the maximum proton kinetic energy behind the target (blue solid line). The laser impinges on the target with p-polarized fields. Classical resonance absorption, also known as the Desinov curve~\cite{Gibbon2005}, is shown as a dashed line.}
    \end{figure}

\begin{figure}
    \centering
    \includegraphics[width=0.49\textwidth]{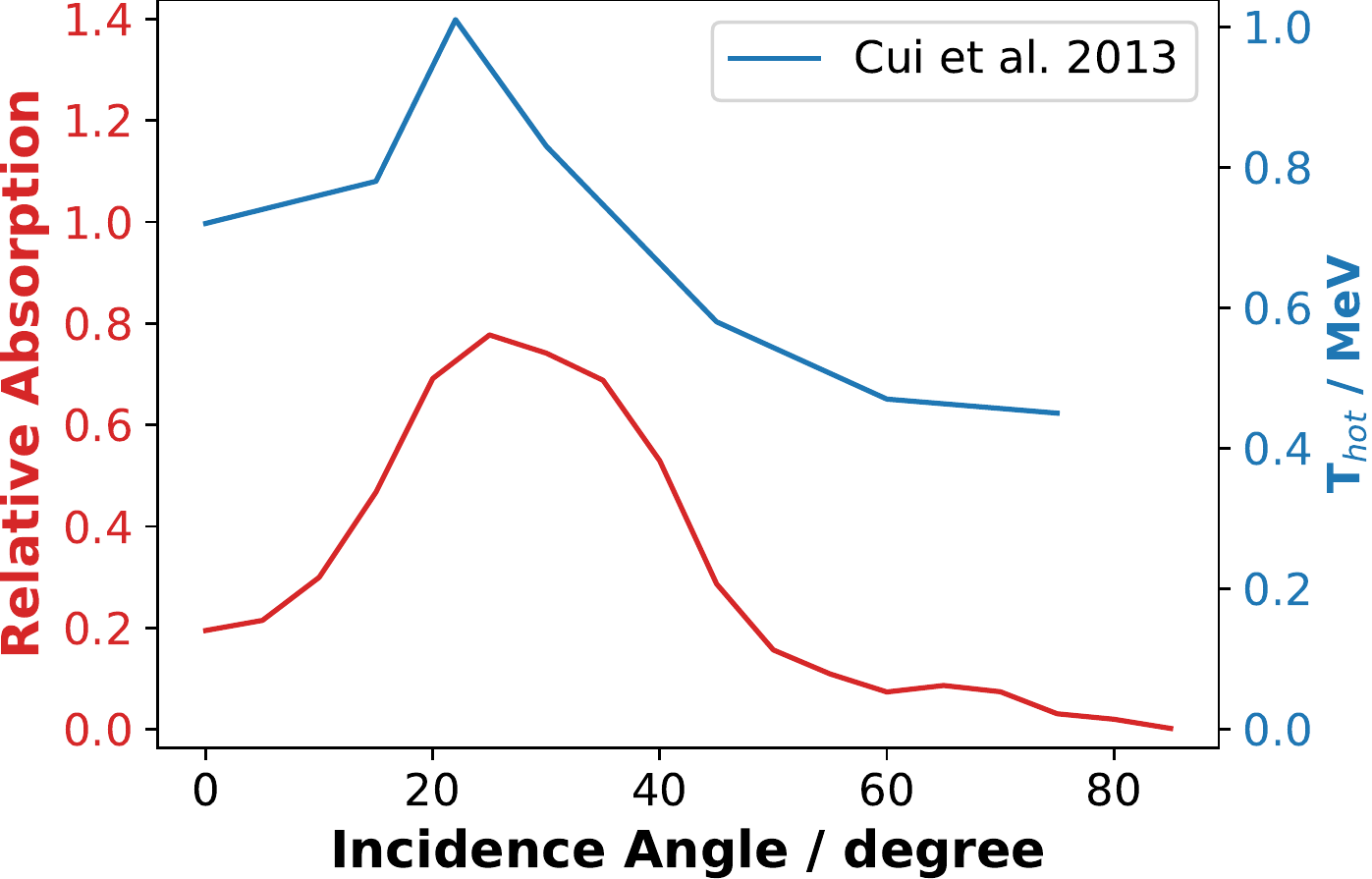}
    \caption{Plot of the absorption of energy in the Lorentz boosted simulation in comparison to the data from Cui et al. \cite{Cui2013}.}
    \label{fig:absVsCui}
\end{figure}

    To illustrate the capabilities of this method, however, the relative laser absorption of a p-polarized laser impinging on a hydrogen plasma target was measured for varying laser incidence angles using the above method in the Smilei PIC code. 
    The resulting absorption curve is shown in \autoref{fig:absscan}. 
    The results agree well with 2D simulations by Cui et al.~\cite{Cui2013} using a similar target and laser (see \autoref{fig:absVsCui} for a comparison).

\subsection*{Explicit Lorentz Boost for oblique Laser Incidence\label{sec:LorentzAppendix}}
In the following section we discuss the full transformation in more detail, and explicitly calculate the relations we mentioned before.
Starting from the transformation matrix in \autoref{eq:lorentzOne} the full derivation will be done for all quantities in the system.

\begin{figure}
    \centering
    \includegraphics[width=0.49\textwidth]{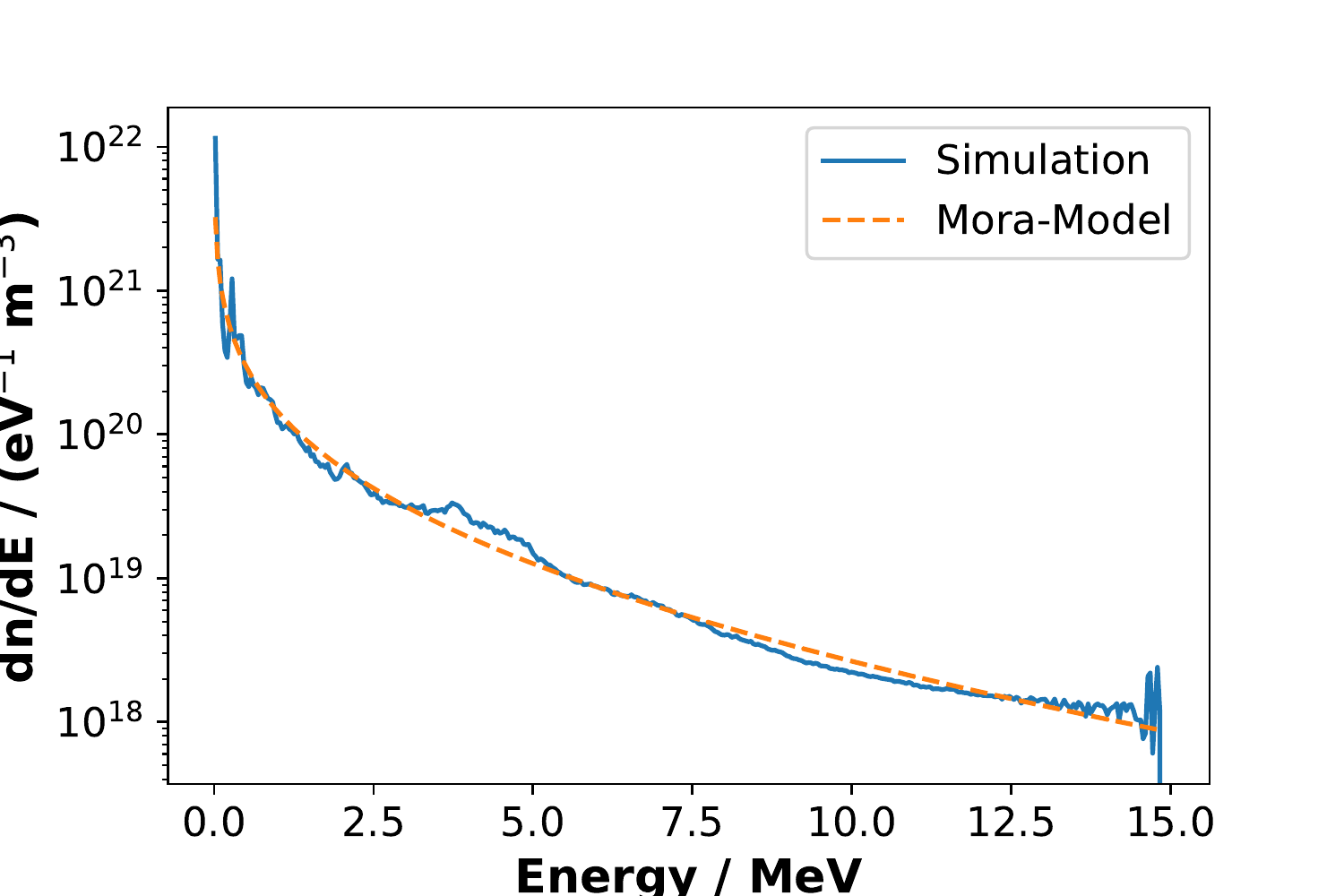}
    \caption{Example result for a Lorentz-boosted simulation with an angle of 40 degrees. The dashed line denotes the fit of Mora's model \cite{Mora2003} to the data.}
    \label{fig:40degreeplot}
\end{figure}

Firstly, the four-position $R$, the four-momentum $P$, the four-wave vector $K$ and the four-current $J$ are given as follows:
	\begin{align}
		\begin{split}
			R &= \left(ct, x, y, z\right)^\top\\
			P &= \left(\gamma  m_0 c, p_x, p_y, p_z\right)^\top\\
			K &= \left(\omega_\tx{L} / c, k \cdot \cos(\theta), k \cdot \sin(\theta), 0\right)^\top\\
			J &= \left(c\rho, j_x, j_y, j_z\right)^\top
		\end{split}    
	\end{align}
where $k=\omega_\tx{L}/c$ is the magnitude of the wave vector, $\rho$ is the charge density and $\vec{j}$ is the current density.
Here, the geometry of the wave vector from \autoref{fig:lorentzscheme} has already been applied, reducing the wave vector to two spatial dimensions. 
By left multiplication of $\Lambda$ these quantities can be transformed into the boosted frame. 
This multiplication yields
\begin{align}
    \begin{split}
        R' &= \left(\gamma_0(ct - y \beta_0), x, \gamma_0(y - ct \beta_0), z\right)^\top\\
        P' &= \left(\gamma_0(\gamma m_0c - p_y \beta_0), p_x,\gamma_0(p_y - \gamma m_0 c \beta_0), p_z \right)^\top\\
        K' &= \left(\omega_\tx{L}/(c\gamma_0), k_0/\gamma_0,0,0 \right)^\top\\
        J' &= \left(\gamma_0(c\rho -j_y\beta_0), j_x, \gamma_0(j_y - c\rho\beta_0), j_z \right)^\top
    \end{split}
\end{align}
where a prime indicates quantities in the transformed system and $\beta_0=v_y/c=\sin(\theta)$. 
Most importantly here we find $k'_y = 0$ and $\omega'_\tx{L}=\omega_\tx{L}/\gamma_0$. 
Also, since the particles are assumed cold at $t=0$, we find for the initial density $\rho'_0 = \gamma_0 \rho_0$.\\
The next transformation is for the electromagnetic fields. 
Here, we differentiate between s- and p-polarized incidence lasers. 
To transform the electric and magnetic fields of the incoming laser, the electromagnetic tensor is used:
	\begin{align}
		F_\tx{s-pol}^{\mu\nu} &= \begin{pmatrix} 0 & 0 & 0 & - E_z/c\\
			0 & 0 & 0 & B_y\\
			0 & 0 & 0 & -B_x\\
			E_z/c & -B_y & B_x & 0
		\end{pmatrix}\\
		F_\tx{p-pol}^{\mu\nu} &= \begin{pmatrix} 0 & -E_x/c & -E_y/c & 0\\
			E_x/c & 0 & -B_z & 0\\
			E_y/c & B_z & 0 & 0\\
			0 & 0 & 0 & 0
		\end{pmatrix} \; .
	\end{align}
The Lorentz transformation of such a tensor is given by:
	\begin{align}
		F^{\mu'\nu'} &= \Lambda^{\mu'}{}_\mu \Lambda^{\nu'}{}_\nu F^{\mu\nu} \; ,
	\end{align}
where a prime again indicates quantities in the transformed system.
The calculated fields are
\begin{align}
    \begin{rcases}
        E'_x &= 0\\
        E'_y &= 0\\
        E'_z &= \gamma_0(E_z - v_yB_x)\\
        B'_x &= \gamma_0 (B_x-E_z v_y/c^2)\\
        B'_y &= B_y\\
        B'_z &= 0
    \end{rcases} \; \tx{s-pol}\\
    \begin{rcases}
        E'_x &= \gamma_0(E_x + v_y B_z)\\
        E'_y &= E_y\\
        E'_z &= 0\\
        B'_x &= 0\\
        B'_y &= 0\\
        B'_z &= \gamma_0(B_z + E_x v_y/c^2)
    \end{rcases} \; \tx{p-pol}
\end{align}
where $E'_x,B'_x \overset{!}{=}0$ since the laser is at normal incidence in the boosted system.
For absorption measurements, it is useful to have a look at the transformation of the Poynting Vector $\vec{S}$. We first define in vacuum
\begin{align}
    \vec{S} &= \frac{1}{\mu_0} \vec{E} \times \vec{B} \; ,
\end{align}
where $\mu_0$ is the vacuum permeability. As an example, we will only present the calculation in the p-polarization case. 
The s-polarization calculation is equivalent. 
We find
\begin{align}
    \vec{S}_\tx{p-pol} &= \frac{1}{\mu_0} \left(E_y B_z,-E_x B_z,0\right)^\top\\
    \Rightarrow \vec{S}'_\tx{p-pol} &= \frac{1}{\mu_0} (E'_y B'_z,\underbrace{-E'_x B'_z}_{=0},0)^\top\\
    &= \frac{1}{\mu_0} \left(E'_y B'_z,0,0\right)^\top \; .
\end{align}
We hence find for the magnitude of the transformed Poynting Vector
\begin{align}
    |\vec{S}'_\tx{p-pol}| &= \frac{1}{\mu_0} \sqrt{E_y^{\prime 2} B_z^{\prime 2}}\\
    &= \frac{1}{\mu_0 c} E_y^{\prime 2} \; ,
\end{align}
since $|\vec{B}| = |\vec{E}|/c$. 
On the other hand, inserting the transformation into $\vec{S}$, we find
\begin{align}
    \vec{S}_\tx{p-pol} &= \frac{1}{\mu_0} \begin{pmatrix}E_y' \gamma_0 \left(B'_z - E'_x \frac{v_y}{c^2} \right)\\
    -\gamma_0 \left(E'_x-v_yB'_z\right) \gamma_0 \left(B'_z - E'_x \frac{v_y}{c^2} \right)\\
    0\end{pmatrix}\\
    &= \frac{1}{\mu_0} \begin{pmatrix} \gamma_0 E'_y B'_z\\
    \gamma_0^2 v_y B_z^{\prime 2}\\
    0\end{pmatrix}
\end{align}
such that for the magnitude we have
\begin{align}
    |\vec{S}_\tx{p-pol}| &= \frac{\gamma_0}{\mu_0} \sqrt{E_y^{\prime 2}B_z^{\prime 2} + \gamma_0^2 v_y^2 B_z^{\prime 4}}\\
    &= \frac{E_y^{\prime 2} \gamma_0}{\mu_0 c} \sqrt{1+\gamma_0^2 \frac{v_y^2}{c^2}} \; .
\end{align}
The term in the square root can be resolved elegantly once we remind ourselves of the definition of $v_y$:
\begin{align}
    1+\gamma_0^2 \frac{v_y^2}{c^2} &= 1 + \frac{1}{\cos^2(\theta)}\frac{c^2 \sin^2(\theta)}{c^2}\\
    &= \frac{1}{\cos^2(\theta)}\\
    &= \gamma_0^2 \; ,
\end{align}
and with that we have
\begin{align}
    |\vec{S}_\tx{p-pol}| &= \frac{E_y^{\prime 2} \gamma_0}{\mu_0 c} \; \gamma_0\\
    &= |\vec{S}'_\tx{p-pol}| \cdot \gamma_0^2 \; .
\end{align}
Next, let us consider the transformed quantities in code units, so as to initialize the particles correctly in the PIC code. 
For the space coordinate, we find
\begin{align}
    \tilde{x}' &= \frac{\omega' x'}{c} = \tilde{x}/\gamma_0 \; ,
\end{align}
while for the time coordinate, since $\tilde{y}' \overset{!}{=} 0$:
\begin{align}
    0 \overset{!}{=} \tilde{y}' &= \frac{\omega'}{c} \gamma_0 (y-ct\beta_0)\\
    &= \tilde{y} - \tilde{t} \tilde{v}_y\\
    \Rightarrow \tilde{t}' &= \omega' \gamma_0 \left(t-\frac{y}{c} \beta_0 \right)\\
    &= \tilde{t} - \tilde{y} \tilde{v}_y\\
    &= \tilde{t} - \tilde{t} \tilde{v}_y^2\\
    &= \tilde{t}/\gamma_0^2
\end{align}
Finally, the critical density transforms as
\begin{align}
    \frac{n'_\tx{c}}{n_\tx{c}} &= \frac{\omega^{\prime 2}}{\omega^2} = \frac{1}{\gamma_0^2} \; ,
\end{align}
such that the initial particle densities in code units become
\begin{align}
    \tilde{n}'_0 &= \frac{n'_0}{n'_\tx{c}} = \tilde{n}_0 \cdot \gamma_0^3 \; .
\end{align}
A verification plot for the Lorentz Boost method is displayed in \autoref{fig:40degreeplot} for irradiation under an oblique angle.


\section{Laser Conversion Efficiency\label{sec:E-Appendix}}
The laser conversion efficiency is an important quantity to characterize particle acceleration and especially laser-plasma acceleration.
In order to retrieve information about the energy in the output spectrum of a TNSA experiment, consider first a spectrum $\total N/\total E$ recorded in multiple energy bins of width $\Delta E$. 
In this case, the number of particles in bin $i$ is given by the bin's height multiplied by its width, i.e.
	\begin{align}
		N_i &= \left(\frac{\total N}{\total E}\right)_i \cdot \Delta E \; .
	\end{align}
Hence, the total energy of the particles within the bin could be approximated by multiplying the particles in the bin by the bin's central energy $E_i$. 
Summing over all bins yields the total energy of the particles
	\begin{align}
		E_\tx{tot} &= \sum_i \left(\frac{\total N}{\total E}\right)_i \cdot \Delta E \cdot E_i \; ,
	\end{align}
which can be generalized in the continuous limit $\Delta E \rightarrow 0$, giving
	\begin{align}
		E_\tx{tot} &= \int_0^\infty \frac{\total N}{\total E} \cdot E \, \total E \; .
	\end{align}
Concretely, adjusting for the output format of the neural network models the total energy is given by
	\begin{align}
		E_\tx{tot} &= V \cdot \int_0^{E_\tx{max}} \exp\left(\ln \left(\frac{\total n}{\total E}\right) \right) \cdot E \, \total E \; ,
	\end{align}
where $\ln \left(\frac{\total n}{\total E}\right)$ and $E_\tx{max}$ are given by the neural network models and $V$ is a unit volume. 
To obtain a measure for the energy conversion efficiency then, the above integral should be weighted by the laser pulse energy $E_\tx{L}$, resulting in the maximization problem shown in Eq.~\eqref{eq:e-conversion}.

\section{Neural Network Training and preparation\label{sec:NetworkTraining}}
In this section, we discuss the chosen parameter ranges for the surrogate models based on neural networks. 

Training surrogate models is a tedious and numerically expensive task. 
This means that we have to be clear about the parameters and data used for the training process. 
We will first focus on the data preparation task, and second on the numerical hyperparameters chosen for our model. 
Both parts are important if we want to create fast converging models.

\subsection{Data preparation}
Neural networks can only be as good as the data used for training them. 
Convergence is important and data, therefore, has to be prepared properly. 
We can only investigate the multi-species effect and subsequent optimizations if we take the full spectrum into account.

\begin{figure}
    \centering
    \includegraphics[width=0.55\textwidth]{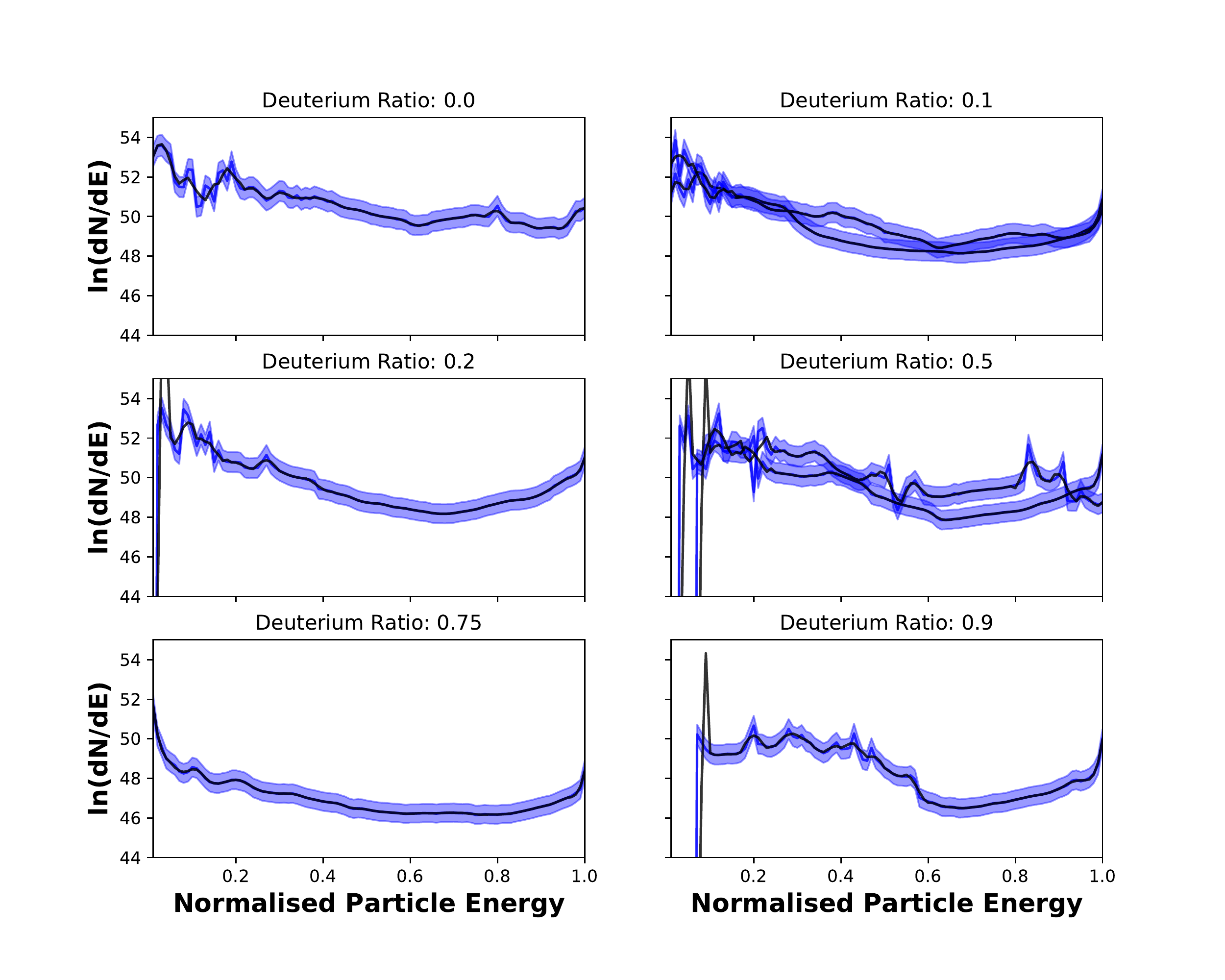}
    \caption{Savitzky-Golay filter with window size 7 and a 3rd order polynomial. Blue indicates the raw data, projected to a wider range for visibility and black indicates the filtered signal.}
    \label{fig:filter}
\end{figure}

The spectral data for the output spectrum is taken on a logarithmic scale since the count rates vary over several orders of magnitude. 
The logarithmic data can directly be used to train a model. 
We tried using the data directly, but convergence was problematic. 
This is due to the noise of the data and the mixture-depending shifts of multi-species plateaus. 
The signal variation in both cases is similar and it is therefore difficult for the network to fit the dependencies. 
To mitigate this we applied a Savitzky-Golay filter \cite{SavitzkySmoothing1964} with a window size of 7 points and a 3rd-order polynomial.
This filter decreased the noise-based fluctuations and allowed subsequent convergence.
We display a comparison for filtered and unfiltered data in \autoref{fig:filter}, which shows, that the major behavior of the curves is reproduced but the bin-to-bin fluctuations in the mid to high energy range are minimized. 

\subsection{Numerical Parameters, Training and Topology}
As none of the architectural parameters for these models were known, some outlying hyperparameters were decided first. For a regression problem, the Rectified Linear Unit (ReLU) activation function is widely used and was added to every layer of the network except the output layer which used the identity activation. Similarly, we chose the mean squared error, suited for regression problems, as loss and it was minimized using the Adam optimizer with $\beta_1 = 0.9$, $\beta_2 = 0.999$ and $\epsilon = \SI{1e-07}{}$. The initial learning rate was 0.001 which was lowered to a minimum of 0.0001 during training should the optimizer detect a plateau in the validation loss value (Keras' \code{ReduceLROnPlateau} feature). In order for the physical parameters to be more manageable numerically, all parameters were divided by the maximum value in their range (see \autoref{tab:physical_parameter_list}) before being given to the model.\\
With these outlying parameters in place, the architecture of the FCNs, i.e. the number of layers and the number of neurons in each layer, was left variable and was optimized for the problem using a hyperparameter tuning method. Keras Tuner allows for extensive hyperparameter tuning using various optimization algorithms~\cite{omalley2019kerastuner}.

Recalling \autoref{sec:HPC}, each simulation output contains information about 100 locations in the energy spectrum of the particles. Hence, for the reduced continuous model, the available data length was $68973 \times 100 = 6897300$ data points. Of these, 81\% were used for training, 9\% were used for validation, and 10\% were used for testing.
Running Keras Tuner on Google Cloud Compute Engine API from a Google Colab Notebook, Bayesian Optimization could be performed for the hyperparameters of the continuous model of hydrogen ions. 
In order to find a model architecture that most accurately describes the simulation results the number of layers and the number of neurons for each layer was first optimized to achieve the lowest possible training loss. 
Every training used a batch size of 256 and an early stopping mechanism. 
After 50 trials, each running training twice in order to lower the chance of a bad local minimum, a suitable architecture was found. However, this optimized model was only tuned to minimize the training loss of the model without considering the validation data at all. To generalize the model, hyperparameter tuning was run again on the optimized architecture, this time with L1 and L2 regularization on each layer as the hyperparameters to be tuned and with the tuning objective set to the mean squared error on the validation set. 
Each hidden layer in the network has L1 regularization strength of \SI{1.98e-06}{} and L2 regularization strength of \SI{3.07e-08}{}. The network achieved a mean squared error of 3.38 on the 620 757 randomly selected validation data points. As a reminder, this number is equal to the mean squared error on the $\ln \left(\frac{\total n}{\total E} (E) \right)$ prediction for input parameters $\{E, \tx{[physical parameters]}\}$.\\
Equivalently, the second model predicting the maximum ion energy could be tuned and optimized. Since the maximum energy is only predicted per simulation and not per energy bin of the energy spectra, the second model was trained on 68 973 unique data points. This significantly smaller dataset made the model training on a home computer feasible.
The optimized model for the maximum energy found L1 regularization strength of \num{2.3e-4} and L2 regularization strength of \num{1.1e-7}.

\bibliography{liquidjetSources}% Produces the bibliography via BibLaTeX.

\end{document}